\documentclass[journal=jacsat,manuscript=article]{achemso}

\usepackage{achemso}
\usepackage{placeins}
\usepackage{graphics}
\usepackage{amssymb,amsfonts}
\usepackage{mathtools}
\usepackage{physics}
\usepackage{subcaption}
\usepackage{graphicx}
\usepackage[table,dvipsnames]{xcolor}
\usepackage{multirow}
\usepackage{caption}
\usepackage{booktabs}
\usepackage{colortbl}
\usepackage{amsmath}
\usepackage{amsopn}
\usepackage{bm}
\usepackage{color}
\usepackage{array}
\usepackage{lscape}
\usepackage{mciteplus}
\usepackage[version=3]{mhchem}
\usepackage[normalem]{ulem}
\usepackage{listings}
\usepackage{enumerate}
\usepackage{lmodern}
\usepackage{mathrsfs}
\usepackage{longtable} 
\usepackage{tabularx} 
\usepackage[symbol]{footmisc} 

\SectionNumbersOn

\newcommand{\cfour}{\textsc{CFOUR}}
\newcommand{\mint}{\textsc{MINT}}
\newcommand{\qcumbre}{\textsc{QCUMBRE}}
\newcommand{\onlinecite}[1]{\hspace{-1 ex} \nocite{#1}\citenum{#1}} 
\newcommand{\siro}{\sigma \rho}
\newcommand{\munu}{\mu \nu}
\newcommand{\numu}{\nu \mu}
\DeclareMathOperator*{\argmax}{arg\,max}

\author{J{\"u}rgen Gauss}
\email{gauss@uni-mainz.de}
\affiliation{Department Chemie, Johannes Gutenberg-Universit{\"a}t Mainz,\\Duesbergweg 10-14, 55128 Mainz, Germany}
\author{Simon Blaschke}
\affiliation{Department Chemie, Johannes Gutenberg-Universit{\"a}t Mainz,\\Duesbergweg 10-14, 55128 Mainz, Germany}
\author{Sophia Burger}
\affiliation{Department Chemie, Johannes Gutenberg-Universit{\"a}t Mainz,\\Duesbergweg 10-14, 55128 Mainz, Germany}
\author{Davide Cianchino}
\affiliation{Fachrichtung Chemie, Universität des Saarlandes, Campus B2.2, D-66123, Saarbr\"ucken, Germany}
\author{Marios-Petros Kitsaras}
\affiliation{Fachrichtung Chemie, Universität des Saarlandes, Campus B2.2, D-66123, Saarbr\"ucken, Germany}
\alsoaffiliation{Laboratoire de Chimie et Physique Quantiques (UMR 5626), Universit{\' e} de Toulouse, CNRS, Bat. 3R1b4, 118 Route de Narbonne, 31062 Toulouse, Cedex 09, France}
\author{Luca Melega}
\affiliation{Dipartimento di Chimica e Chimica Industriale, Universit{\`a} di Pisa, Via G. Moruzzi 13, 56124, Pisa, Italy}
\author{Tommaso Nottoli}
\affiliation{Dipartimento di Chimica e Chimica Industriale, Universit{\`a} di Pisa, Via G. Moruzzi 13, 56124, Pisa, Italy}
\author{Jeremias Oswald}
\affiliation{Fachrichtung Chemie, Universität des Saarlandes, Campus B2.2, D-66123, Saarbr\"ucken, Germany}
\author{Tereza Uhl\'i\v{r}ov\'a}
\affiliation{Department Chemie, Johannes Gutenberg-Universit{\"a}t Mainz,\\Duesbergweg 10-14, 55128 Mainz, Germany}
\author{Chaoqun Zhang}
\affiliation{Department of Chemistry, Yale University,\\New Haven, CT 06520-8107, USA}
\author{Lan Cheng}
\affiliation{Department of Chemistry, Johns Hopkins University,\\Baltimore, MD 21218, USA}
\email{lcheng24@jhu.edu}
\author{Stella Stopkowicz}
\affiliation{Fachrichtung Chemie, Universität des Saarlandes, Campus B2.2, D-66123, Saarbr\"ucken, Germany}
\alsoaffiliation{Hylleraas Centre for Quantum Molecular Sciences, Department of Chemistry, University of Oslo, P.O. Box, Blindern, 1033, N-0315 
Oslo, Norway}
\email{stella.stopkowicz@uni-saarland.de}
\author{Filippo Lipparini}
\affiliation{Dipartimento di Chimica e Chimica Industriale, Universit{\`a} di Pisa, Via G. Moruzzi 13, 56124, Pisa, Italy}
\email{filippo.lipparini@unipi.it}

\title{Use of Cholesky decomposition in the \cfour\ program package}

\begin{document}


\newpage 

\begin{abstract}
An overview is given about our efforts to speed up high-level computations of molecular energies and properties using Cholesky decomposition (CD). We describe the corresponding developments in the \cfour\ quantum-chemical package with a focus on (a) the use of CD in complete-active space self-consistent-field (CASSCF) and coupled-cluster (CC) computations, (b) the implementation of analytic CC gradients using CD, (c) the efficient calculation of magnetic properties (NMR shieldings and magnetizabilities) using CD when carried out with explicit magnetic-field dependent basis functions, i.e., the so-called gauge-including atomic orbitals (GIAOs), (d) the efficient computation of CASSCF response properties using CD, (e) the use of CD in relativistic two- and four-component computations, as well as (f)  computations for molecules in finite magnetic fields using GIAOs together with CD. An outlook on future developments concerning the \cfour\ package is given.
\end{abstract}

\section{Introduction}
The \cfour\ program package\cite{Matthews20c,cfour} focuses on high-accuracy computations of energies and properties for atoms and small to medium-sized molecules using many-body methods.\cite{Shavitt09} Its main features have been recently described in a review article.\cite{Matthews20c} However, the applicability of the highly accurate methods available in \cfour\ is severely hampered by their high computational cost, i.e., the high computational timings and the large demands of core and external memory. This article describes our recent efforts to alleviate this problem and to extend the applicability of the quantum-chemical approaches available in \cfour\ to larger molecules.

To cope with the challenges due to excessive demands of external memory, different strategies have been proposed in the literature.\cite{Almloef82,Eichkorn95,Koch03}
The so-called integral-direct strategies are characterized by a recomputation of the two-electron integrals (in the atomic-orbital (AO) basis) whenever needed and render their storage unnecessary. First applied to Hartree-Fock (HF) calculations\cite{Almloef82,Cremer85,Haeser91,Horn91,Haeser92b,Weiss93}, this strategy has been later also used in second-order M{\o}ller-Plesset perturbation (MP2) theory\cite{Moller34,Cremer00} energy\cite{Saebo89,Haeser89,Haeser91c} and gradient computations\cite{Haase93,Frisch90,Frisch90a} as well coupled-cluster (CC)\cite{Shavitt09} calculations.\cite{Koch96,Halkier97a,Schuetz99} The increase in the computational timings due to the repeated evaluation of the two-electron integrals can be 
reduced by efficient integral prescreening\cite{Haeser91} or the use of semi-direct techniques.\cite{Haeser91,Frisch90a}

Another important step towards a reduction of the computational timings has been the introduction of resolution-of-identity (RI), also referred to as density-fitting (DF) techniques in electron-correlated computations. These techniques employ an additional expansion of the density matrix in terms of an auxiliary basis set and thus reduce the 
time-consuming computation of the two-electron integrals to the evaluation of three-center integrals. This idea was first pursued in density-functional theory\cite{Whitten73,Baerends73,Dunlap79} and then later exploited as well in conventional quantum-chemical methods. To be mentioned is here the pioneering work by Vahtras {\it et al.}\cite{Vahtras93}
and the work by Eichkorn {et al.}\cite{Eichkorn95} which put RI/DF on a solid basis and provided optimized auxiliary basis sets for the RI/DF procedure. RI implementations of MP2 energies\cite{Feyereisen93}and gradients\cite{Weigend97} were subsequently reported. The use of RI/DF techniques is nowadays standard and many quantum-chemical programs apply them to enable and speed up large-scale computations.

An at the first sight unrelated idea to speed up quantum-chemical computations was put forth in 1977 by Beebe and Linderberg,\cite{Beebe77} when they suggested to use Cholesky decomposition (CD)\cite{Benoit24} to compress the two-electron integrals. It took a while before this idea was proven useful: Koch {\it et al.}\cite{Koch03} demonstrated in 2003 how CD can be efficiently implemented. Their paper led to an increased interest in the use of CD in quantum-chemical calculations. Noteworthy are here the efforts in the Lund group\cite{Aquilante09c} to speed up multi-configurational computations\cite{Aquilante08a,Aquilante08b} and by Pito{\v n}ák {\it et al.}\cite{Pitonak11}, Epifanovsky {\it et al.}\cite{Epifanovsky13}, and Bozkaya\cite{Bozkaya14,Bozkaya16,Bozkaya16a} to exploit CD in the framework of CC (and equation-of-motion (EOM)-CC) computations. CD exploits that the two-electron integrals can be considered as elements of a symmetric (Hermitian) positive semidefinite matrix and thus represented as a product of two triangular matrices. Such a matrix, for large enough molecules, has a low rank, which translates into the fact that not all columns (rows) of the triangular matrix are needed to represent the two-electron integrals with sufficient accuracy. The latter fact
can be exploited to achieve substantial savings in computation times as well in memory demands. Furthermore, the decomposition can be carried out without first constructing the full two-electron integral matrix. For reviews on the use of CD in quantum chemistry, see Refs.~\onlinecite{Aquilante11} and \onlinecite{Pedersen23}.

An important advantage of CD over RI/DF is that the CD of the two-electron integrals provides upper bounds for the accuracy of the approximated integrals. However, unlike RI/DF, integral evaluation in CD is not restricted to three-center integrals. A restriction to three-center integrals is nevertheless possible by invoking additional approximations (one-center CD or atomic CD, see Ref.~\onlinecite{Aquilante07a}) which, however, spoil the upper-bound feature of CD. Conceptually, it has also been essential to realize the close relationship between CD and RI/DF\cite{Aquilante08c,Pedersen09} in the way that CD is DF with an on-the-fly constructed auxiliary basis, the so-called Cholesky basis. This observation turned out to be fruitful for the formulation of analytic gradients with Cholesky decomposed integrals,\cite{Aquilante08c} as the decomposition can be only applied to the undifferentiated integrals and not the differentiated ones. However, up to date several CD based implementations of CD based analytic gradients\cite{Aquilante08c,Bostroem14,Delcey14,Feng19,Schnack22,Melega26} have been reported and are in use.

Motivated by the work of Krylov and coworkers\cite{Epifanovsky13} on CC and EOM-CC methods using CD and due to the involvement of one of the authors of \cfour\ in the formulation and implementation of CD-CC and CD-EOM-CC gradients within the {\sc Q-Chem} package,\cite{Feng19} we started to implement CD based techniques in the \cfour\ package with a focus on the computation of properties. After some experimenting,\cite{Hilgenberg17} a solid CD procedure was implemented within the new integral package {\mint}\cite{mint} and used for HF, CASSCF,\cite{Roos80} as well as CC singles and doubles (CCSD)\cite{Purvis82} computations. An important ingredient in \cfour\ has been straight from the beginning the exploitation of Abelian point-group symmetry. We also adopted advances by others concerning the CD procedure, in particular the two-step decomposition scheme as suggested by Aquilante {\it et al.}\cite{Aquilante11} and implemented efficiently by Folkestad {\it et al.}\cite{Folkestad19} as well as Zhang {\it et al.}\cite{Zhang21} New developments were made concerning the use of CD in the computation of magnetic properties such as NMR shieldings and magnetizabilities, the use of CD in finite-field (ff) computations, and the use of CD based two-electron integrals in relativistic computations.

In the following we will provide a status report on the use of CD within the \cfour\ package. We start with a review on CD applied to the two-electron integrals (Section \ref{section2}), describe the corresponding implementation of CD in \cfour\ (Section \ref{section3}). This is followed by discussions how Cholesky decomposed two-electron integrals are used in HF, complete active space self-consistent field (CASSCF),\cite{Roos80} and CCSD computations (Sections \ref{section4} and \ref{section5}). The calculation of molecular properties using CD is described in Section \ref{section6} concerning analytically evaluated nuclear forces, in Section \ref{section7} concerning the computation of magnetic properties, and in section \ref{section8} concerning linear-response properties computed at the CASSCF level. Section \ref{section9} deals with the use of CD in relativistic calculations discussing the use of CD in full four-component computations as well as in two-component computations with inclusion of spin-orbit coupling. Section \ref{section10} is devoted to the use of CD in ff calculations for the treatment of molecules in external magnetic fields. Our status report concludes with an outlook on future developments in the \cfour\ package concerning cost-reducing strategies for computations on larger molecules.

\section{Cholesky decomposition}
\label{section2}
Given a symmetric, positive definite matrix $\bf A$ its Cholesky decomposition (CD) is the factorization
\[
{\bf A }= {\bf LL}^T,
\]
where $\bf L$ is a lower triangular matrix with positive diagonal elements. This factorization is unique and provides a numerically stable representation of $\bf A$ while reducing both storage requirements and computational cost of many linear algebra operations\cite{Allaire08}. If $\bf A$ is only positive semidefinite, the standard Cholesky factorization may break down because zero pivots can occur.
Pivoted Cholesky decomposition can instead be used to identify and discard redundant information, yielding a compact factorization whose rank is determined by the decomposition threshold.
The properties of CD for positive semidefinite matrices form the basis for its application to quantum chemistry, where the matrix to be decomposed is the electron-repulsion integral (ERI) matrix $(\mu\nu|\rho\sigma)$, with composite indices $\mu\nu$ and $\rho\sigma$. Such a matrix is well known to be numerically rank deficient, and thus pivoted CD, as originally introduced in this context by Beebe and Linderberg,\cite{Beebe77} provides a natural route to a compact, nonredundant representation of the ERIs. 

\begin{figure} [H]
    \centering
    \includegraphics[width=8cm]{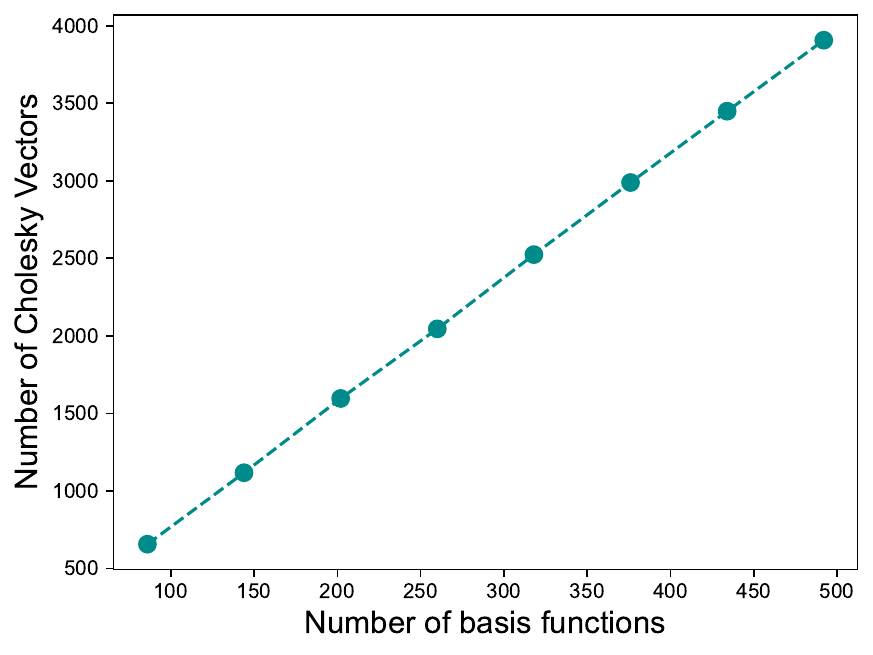}
    \caption{Number of CVs for the homologous series of straight-chained alkanes $\mathrm{C}_2\mathrm{H}_{2\mathrm{n}+2}$ in a cc-pVTZ basis\cite{Dunning89} using a Cholesky threshold $\tau$ of $10^{-5}$.} 
    \label{fig:PlotCV-BF}
\end{figure}

The algorithm for fully-pivoted CD is remarkably simple. Let
\[
D^{1}_{\mu\nu} = (\mu\nu|\mu\nu)
\]
be the diagonal of the ERI tensor. Let the first pivot $(\bar{\mu}\bar{\nu})_1$ be the pair of indices for which the diagonal is maximum, i.e., 
\[
(\bar{\mu}\bar{\nu})_1 = \argmax_{\mu\nu} D^{1}_{\mu\nu}
\]
The first Cholesky vector (CV) $L^1$ is then defined as
\[
L^1_{\mu\nu} = \frac{(\mu\nu|(\bar{\mu}\bar{\nu})_1)}{\sqrt{D^1_{(\bar{\mu}\bar{\nu})_1}}}
\]
All other CVs can be obtained through the following recursion:
\begin{equation}
\label{eq:diag}
D^{K+1}_{\mu\nu} = D^{K}_{\mu\nu} - (L^K_{\mu\nu})^2
\end{equation}
\begin{equation}
(\bar{\mu}\bar{\nu})_{K+1} = \argmax_{\mu\nu} D^{K+1}_{\mu\nu}
\end{equation}
\begin{equation}
\label{eq:subtr}
L^{K+1}_{\mu\nu} = \frac{1}{\sqrt{D^{K+1}_{(\bar{\mu}\bar{\nu})_{K+1}}}} \left [ (\mu\nu|(\bar{\mu}\bar{\nu})_{K+1}) - \sum_{J=1}^K L_{\mu\nu}^JL^J_{(\bar{\mu}\bar{\nu})_{K+1}} \right  ]
\end{equation}
The list of pivots $\{ (\bar{\mu}\bar{\nu})_{K}\}_{K=1}^{M}$, commonly denoted simply by $\{ |K)\}_{K=1}^{M}$, is referred to as the \emph{Cholesky basis}.
The decomposition is truncated when 
\[
\max_{\mu\nu} D^{K+1}_{\mu\nu} < \tau,
\]
where $\tau$ is a user-defined threshold. By the Cauchy-Schwarz inequality, this guarantees that the absolute error in every reconstructed ERI is bounded by the same threshold $\tau$\cite{Aquilante11}. Thus, CD provides not only a compact representation of the ERIs, but also one whose accuracy is rigorously controlled by a single user-defined parameter. To better illustrate the compactness of the representation, we report in Fig.~\ref{fig:PlotCV-BF} the number of CVs obtained for a series of increasingly large straight-chained alkanes as a function of the number of basis functions. A virtually perfect linear relation is easily observed, demonstrating the effectiveness of CD in compressing the ERIs information.

\section{Implementation of Cholesky decomposition in \cfour}
\label{section3}
The first production implementation of CD in \cfour\ was developed within the Mainz Integral Package (\mint)\cite{mint} in 2017 and initially employed for exploratory self-consistent-field calculations.\cite{Matthews20c,Nottoli21b} The implementation adopted the partially pivoted algorithm proposed by Koch {\it et al.}\cite{Koch03} and was fully symmetry adapted.
Although the implementation made extensive use of efficient level 2 and level 3 BLAS routines, the subtraction step\cite{Koch03} (Eq.~\ref{eq:subtr}) remained computationally expensive, ultimately dominating the overall computational cost of the procedure and limiting its practical use. 
In 2019, Folkestad {\it et al.}\cite{Folkestad19} introduced a new algorithm to compute the CD based on a two-step procedure, as originally proposed by Aquilante.{\it et al.}\cite{Aquilante11} 
The two-step algorithm is based on two main ideas. The most important one is the formal equivalence between CD and RI/DF.\cite{Aquilante08c,Pedersen09} 
By inserting the resolution of the identity with respect to a non-orthogonal fitting basis $\{|P)\}_{P=1}^{N_{\rm f}}$ of dimension $N_{\rm f}$, one gets
\begin{equation}
    \label{eq:DF1}
    (\mu\nu|\rho\sigma) \approx \sum_{PQ} (\mu\nu|P)S_{PQ}^{-1}(Q|\rho\sigma)
\end{equation}
where $S_{PQ} = (P|Q)$ is the metric for the fitting basis. Since the fitting metric is symmetric positive (semi)definite, it admits a factorization of the form $S = MM^T$. Substituting:
\begin{equation}
    \label{eq:DF}
    (\mu\nu|\rho\sigma) \approx \sum_K\sum_{PQ} (\mu\nu|P)M_{PK}^{-T}M_{KQ}^{-1}(Q|\rho\sigma)
\end{equation}
where we denote with the superscript $-T$ the inverse transpose matrix.
CD can therefore be interpreted as a special case of DF, where the fitting basis $\{|P)\}$ is the Cholesky basis $\{ (\bar{\mu}\bar{\nu})_{P}\}$ defined before, and where the CVs are computed by solving the triangular linear system
\begin{equation}
    \label{eq:CVsLT}
    L^K_{\mu\nu} = \sum_{P} (\mu\nu|P) M^{-T}_{PK}.
\end{equation}
Since the metric associated with the Cholesky basis is simply a small principal submatrix of the ERI matrix, one can easily compute the CVs by first using an efficient dense linear-algebra routine (e.g., LAPACK's DPOTRF\cite{lapack}) to compute the metric's Cholesky factorization, and then use the appropriate lower triangular solver (e.g., DTRSM) to compute the CVs. 
Although this reformulation does not reduce the formal computational scaling with respect to the subtraction in Eq.~\ref{eq:subtr}, it recasts the most expensive operations in terms of highly optimized dense linear-algebra routines, thereby greatly improving computational efficiency.

The second key idea concerns the efficient construction of the Cholesky basis. It relies on the observation that each updated diagonal element $D^{(K)}$, defined in Eq.~\ref{eq:diag}, is non-increasing with $K$. 
Therefore, once a diagonal element falls below the decomposition threshold, the corresponding row and column can be permanently discarded. The CD can thus be performed on a progressively reduced matrix, whose dimension decreases monotonically throughout the procedure, leading to a substantial reduction in computational cost.

\begin{figure} [H]
    \centering
    \includegraphics[width=15cm]{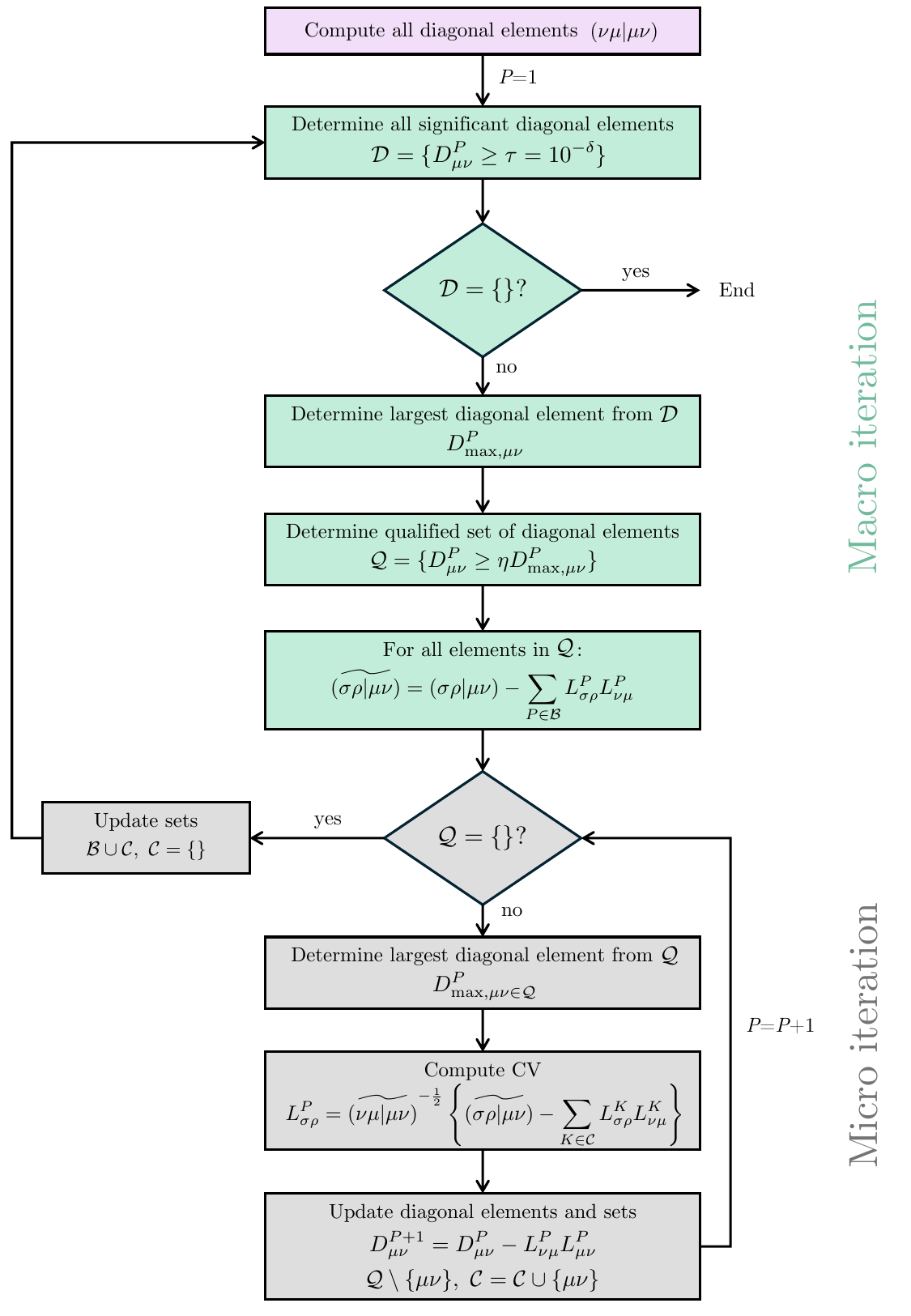}
    \caption{First step of the two-step algorithm for computing the CD of the two-electron integrals in {\mint}.}
    \label{fig:Two-Stepscheme}
\end{figure}

In \cfour, we have implemented the two-step procedure in the variant proposed by Zhang {\it et al.}\cite{Zhang21}. In Zhang's algorithm, once an integral quartet has been computed, contributions involving discarded indices—i.e., rows and columns whose diagonal elements have fallen below the decomposition threshold—are removed, while the remaining integrals are retained in memory for use in subsequent iterations. This avoids their repeated evaluation. The additional memory requirement remains modest because the number of significant diagonal elements grows only linearly with the basis-set size in the asymptotic limit.
Furthermore, the implementation is fully symmetry adapted, as we work with symmetry-adapted linear combinations (SALC) of atomic orbitals. Specifically, only those elements of $L^K_{\mu\nu}$ for which $\Gamma_\mu \otimes \Gamma_\nu = \Gamma_K$, where $\Gamma_\mu$ is the irreducible representation (irrep) of the $\mu$-th basis function and $\Gamma_K$ 
denotes the irrep of the corresponding CV, are computed. We illustrate the implementation of the first step of the CD with a flowchart in Fig.~\ref{fig:Two-Stepscheme}. All the quantities that appear in the flowchart are defined in Table \ref{tab:Two-Step}.

\begin{table} [H]
    \centering 
    \caption{The various sets used in the two-step CD algorithm.}
    \begin{tabular}{ll}
        \toprule
        Set & Content \\
        \midrule
        $\mathcal{B}$ & Chosen Cholesky basis functions \\
        $\mathcal{D}$ & All diagonal elements greater than the Cholesky threshold $\tau$\\
        $\mathcal{Q}$ & All qualified diagonal elements ($>\eta D_{\mathrm{max}}^P$, $\eta$: span factor) \\
        $\mathcal{C}$ & Indices for which the corresponding CV was already computed in the micro iteration\\
        \bottomrule 
    \end{tabular}
    \label{tab:Two-Step}
\end{table}
An important remark is that if perturbed vectors need to be computed as well, as for the calculation of magnetic response properties (see section \ref{sec:magnetic}), this can be done directly during the second step, as the Cholesky basis is the same for both standard and perturbed vectors. 

\begin{figure}
    \centering
    \includegraphics[width=0.8\linewidth]{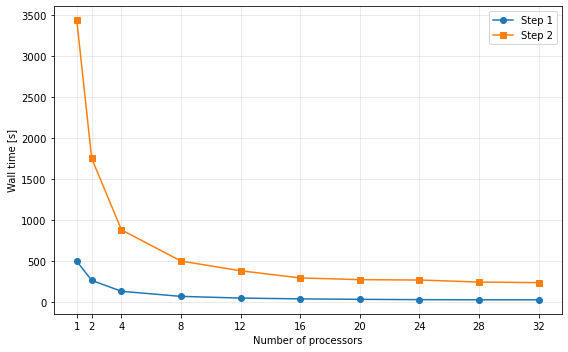}
    \includegraphics[width=0.8\linewidth]{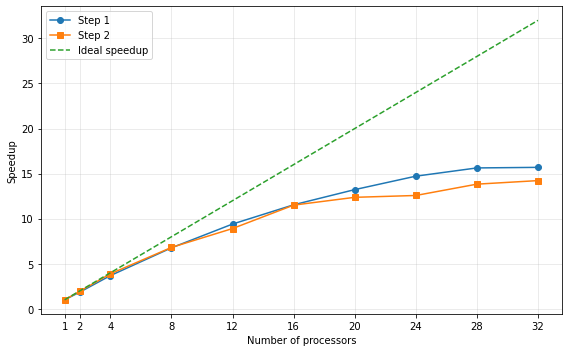}
    \caption{Elapsed time (top panel) and parallel speedup (bottom panel) for the first and second step of the two-step CD determination of both CVs and magnetically perturbed CVs (with GIAOs) for coronene (D$_{2h}$ symmetry, tz2p basis set (684 basis functions)). Calculations were performed on a computer node equipped with two 32 cores Intel Xeon Gold 6438Y+ CPUs running at 2.0 GHz (4.0GHz turbo boost) and 512 GB of DDR5 RAM.}
    \label{fig:scaling}
\end{figure}

The implementation in \cfour\ has been parallelized using shared-memory \textsc{OpenMP} directives. The main source of computational cost stems from the calculation of the (perturbed) integrals, as expected. To assess the performance of the implementation, we compute the CD, including both standard and magnetic perturbed CVs, for coronene enforcing $D_{2h}$ point-group symmetry and employing the Karlsruhe tz2p basis set\cite{Schaefer92} (684 basis functions). The results are shown in Fig.~\ref{fig:scaling}. The calculations were performed on a computer node equipped with two 32 cores Intel Xeon Gold 6438Y+ CPUs running at 2.0 GHz (4.0 GHz turbo boost) and 512 GB of DDR5 random access memory (RAM). The top panel reports the total elapsed (wall) time as a function of the number of cores used for the calculation for both the first (blue line) and second (orange line) steps. Note that the second step includes the calculations of three sets of perturbed vectors, on top of the standard CVs. Both the first and second step show good parallel scaling up to 16 cores, then the performance stagnates, likely due to memory bandwidth saturation. Nevertheless, the whole calculation can be performed in less than 3 minutes using 24 cores. 

\section{Quantum-chemical methods using Cholesky decomposition}
\subsection{Self-consistent field methods: Hartree-Fock (HF) and complete active space self-consistent field (CASSCF) using Cholesky decomposition}
\label{section4}
The use of the CD is particularly advantageous for self-consistent field (SCF) methods. In a conventional (i.e., non-integral-direct) SCF calculation, the leading computational cost (without integral prescreening) scales as $\mathcal{O}(N^4)$ due to the construction of the Fock matrix, where $N$ is the number of basis functions,  When CD is employed, one can show that the evaluation of the Coulomb contribution to the Fock matrix scales as $\mathcal{O}(N^2M)$, while that of the exchange contribution scales as $\mathcal{O}(ON^2M)$, where $M$ is the number of Cholesky vectors and $O$ is the number of occupied orbitals. Consequently, CD provides a significant reduction in computational cost. Furthermore, these operations can be implemented efficiently using level 2 and level 3 BLAS routines and are readily parallelized.
In \cfour, CD is available for restricted (RHF), unrestricted (UHF), and high-spin restricted open-shell Hartree--Fock (ROHF) calculations.\cite{Nottoli21b} The resulting SCF equations can be solved using either the conventional SCF procedure or a robust quadratically convergent algorithm based on the norm-extended optimization (NEO) framework\cite{Jorgensen83,Jensen86}.

For complete active space self-consistent field (CASSCF) calculations,\cite{Roos80} CD substantially reduces the computational cost of the orbital-optimization step, primarily by reducing the cost of the AO-to-MO integral transformation. In fact, for active spaces up to CAS(12,12) to CAS(14,14), the computational cost of a CASSCF calculation is generally dominated by the AO-to-MO transformation of the two-electron integrals, which requires in general $\mathcal{O}((O+A)N^4)$ floating-point operations with $A$ as the number of active orbitals and with $O$ in the context of CASSCF referring to the number of internal orbitals. In our implementation, the Cholesky vectors are fully transformed to the MO basis. This transformation requires $\mathcal{O}(N^3M)$ operations and formally corresponds to quartic scaling (since $M$ grows approximately linearly with $N$). In practice, the effective prefactor is much smaller than $(O+A)$, thus enabling significant cost savings. In addition, the inactive and active Fock matrices, as well as the $Q$ matrix, must be constructed. The leading computational costs for these operations are $\mathcal{O}(N^2AM)$, $\mathcal{O}(N^2OM)$, and $\mathcal{O}(A^4M)$, respectively. For the quadratically convergent implementation, where the product between the MO Hessian and a trial vector is required, so-called one-index-transformed Fock matrices must also be evaluated, and their construction scales as $\mathcal{O}(N^2AM)$ and $\mathcal{O}(N^2OM)$. Therefore, all these operations remain computationally less demanding than the AO-to-MO integral transformation step. Further details on the implementation can be found in Ref.~\citenum{Nottoli21}.

\begin{figure}
    \centering
    \includegraphics[width=0.9\textwidth]{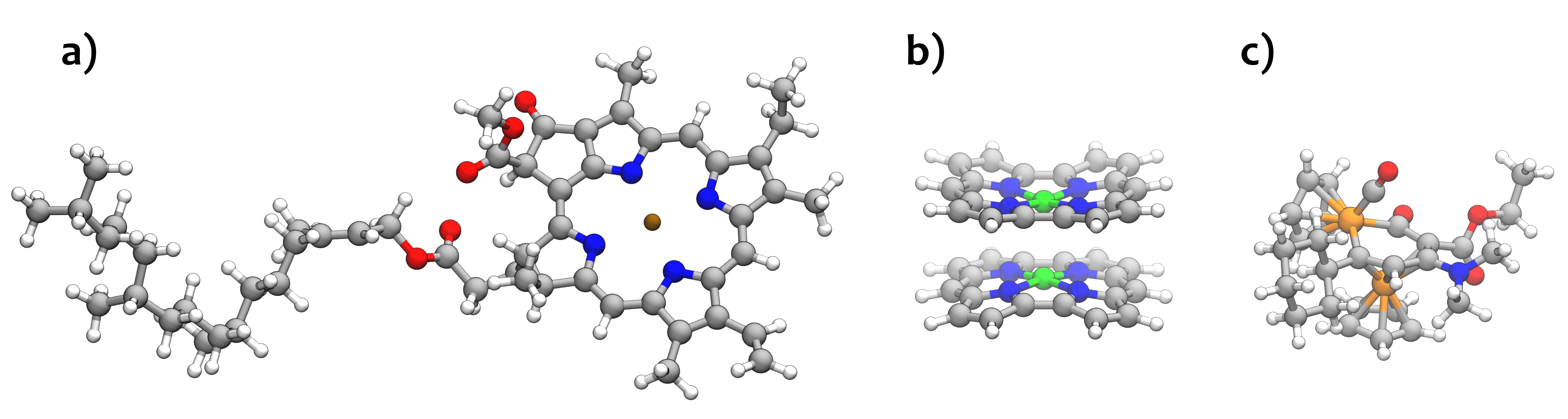}
    \caption{Molecular representation of the chlorophyll molecule (a), face-to-face stacked Ni(II) norcorrole dimer (b), ferrabenzene (c). Carbon is represented in silver, hydrogen in white, oxygen in red, nitrogen in blue, nickel in green, magnesium in brown, and iron in orange.}
    \label{fig:cdcas_systems}
\end{figure}

An additional advantage of employing CD is the reduced memory requirement. To achieve optimal performance, the algorithm stores the complete set of Cholesky vectors in core memory, requiring up to $MN(N+1)/2$ words of storage when no point-group symmetry is present. In addition, the implementation requires several auxiliary arrays of size $N^2$ together with thread-private copies required for OpenMP parallelization.
\begin{figure}[h!]
    \centering
    \includegraphics[width=0.7\textwidth]{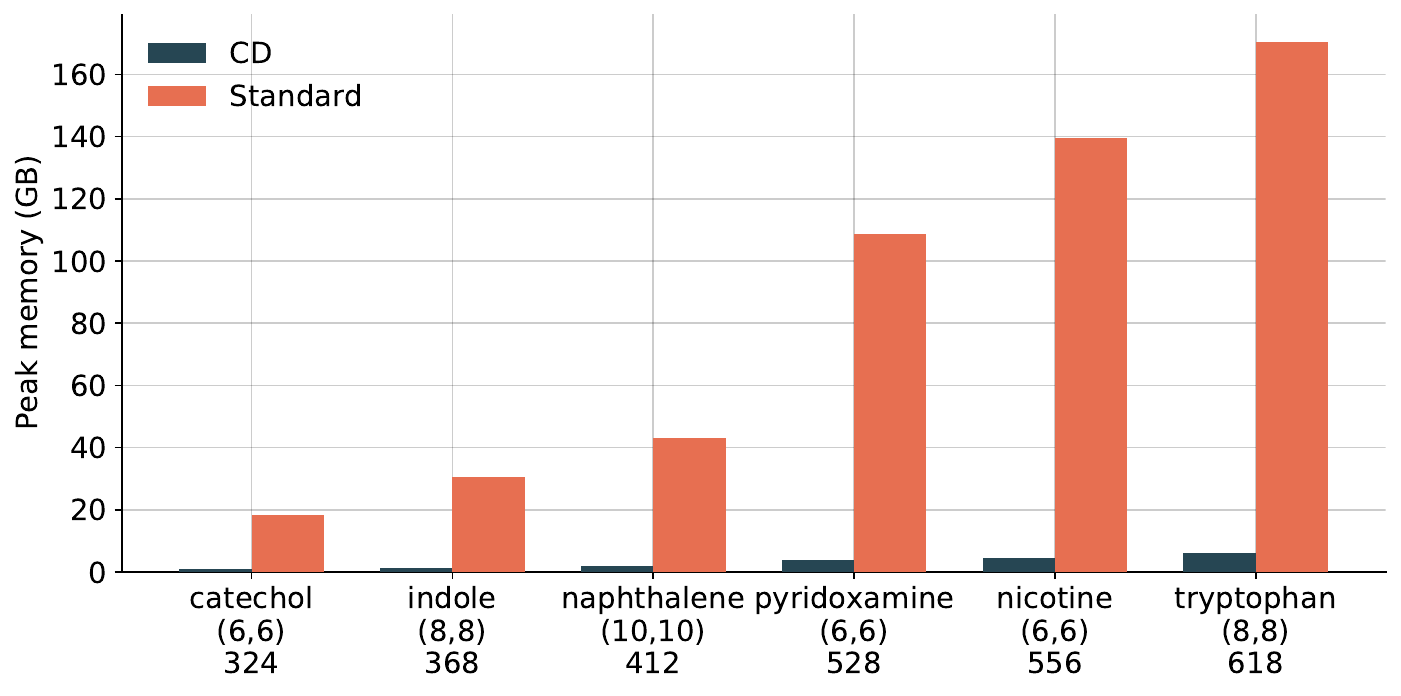}
    \caption{Peak memory consumption for a single-point CASSCF calculation using the cc-pVTZ basis\cite{Dunning89} set with (CD) and without CD (Standard). The calculations do not exploit point-group symmetry. Below each molecule name, the corresponding active space and number of basis functions are reported.}
    \label{fig:mem_CAS}
\end{figure}
The benefits of the CD approach are demonstrated through calculations performed for a selection of molecular systems using the cc-pVTZ basis set,\cite{Dunning89} without enforcing point-group symmetry, ranging from 324 to 618 basis functions. A Cholesky decomposition threshold $\tau$ equal to $10^{-4}$ was used. In Fig.~\ref{fig:mem_CAS}, we report the peak memory consumption for these systems, highlighting the substantial reduction achieved by employing CD. The largest memory requirement is approximately 6~GB, for the tryptophan molecule. The total elapsed wall time required for the six calculations is less than five minutes using 28 threads of an Intel {Xeon Gold 6140M} CPUs, running at 2.30~GHz, further demonstrating the efficiency of the implementation. In general, the present implementation enables routine full in-core calculations for systems containing about 1000 basis functions using modest computational resources. Nevertheless, we also demonstrate that larger calculations are feasible, as shown by a CAS(12,12)/cc-pVTZ calculation\cite{Nottoli21} on the chlorophyll molecule (shown in Fig.~\ref{fig:cdcas_systems}a), comprising 2962 basis functions. The calculation required 12 iterations and took approximately 12~h 15~min to converge to 10$^{-7}E_h$ in the root-mean square (RMS) CASSCF gradient.

In addition, the present CD-CASSCF code has been used to study transition-metal systems. For example, it has been employed to investigate the aromaticity of a Ni(II) norcorrole system and its face-to-face stacked dimer (shown in Fig.~\ref{fig:cdcas_systems}b) through the calculation of magnetically induced current densities,\cite{Juselius04,Sundholm16} providing insights into the role of near-degeneracy effects at the metal center.\cite{Wang24} As a benchmark, we report the timing for the CASSCF(16,15) wavefunction optimization for the dimer using the def2-SVP basis set\cite{Weigend05} (778 basis functions) and a CD threshold $\tau$ of 10$^{-5}$. The calculation was performed on an AMD EPYC 7282 processor using 32 threads. It required 21 iterations and approximately 10~h 40~min to converge, reaching an RMS gradient of 10$^{-10}E_h$. We note that in this case the rate-determining step in the calculation is given by the full configuration interaction (FCI) part of the code, which involves over 41 million Slater determinants. In another study, the code has been used to characterize the geometry as well as the aromatic character of a newly synthesized ferrabenzene complex,\cite{Benetti25} whose molecular representation is shown in Fig.~\ref{fig:cdcas_systems}. The calculation was performed on the same AMD EPYC 7282 processor using 32 threads and the def2-SVP basis set (675 basis functions), with a (14,14) active space and a 10$^{-5}$ CD threshold. The CASSCF wave function optimization converged to 10$^{-7}E_h$ in the RMS of the CASSCF gradient in 21 iterations and took about 1~h 30~min.

In addition, CD techniques have been employed in
state-averaged CASSCF calculations using the Super-CI algorithm in combination with an implicit solvent treatment.\cite{Schank25}\\
Future developments could exploit scalable FCI approaches such as many-body expanded FCI (MBE-FCI),\cite{Eriksen21} density matrix renormalization group (DMRG),\cite{Chan11,Schollwoeck05} or general selected CI approaches\cite{Huron73,Tubman16,Schriber16,Liu16,Sharma17} to extend the active space accessible within CASSCF, promising more systematic treatments of challenging electronic structures. This has already been demonstrated by recent implementations of MBE-CASSCF\cite{Greiner24a} and DMRG-CASSCF,\cite{Gianni26} enabling calculations with active spaces substantially larger than those accessible by conventional FCI solvers. 

\subsection{Coupled-cluster methods using Cholesky decomposition}
\label{section5}
\begin{figure}
    \centering
    \includegraphics[width=0.5\linewidth]{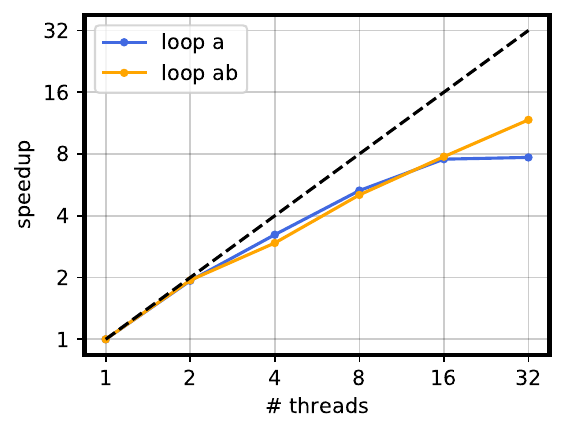}
    \caption{Speedup plot for CD based CCSD/cc-pVTZ calculations on coronene. For both algorithms, we plot the ratio between iteration time for the serial calculation and the iteration time of the parallel ones. Both axes are in the log2 scale. Reproduced from Ref.~\citenum{Nottoli23}, with the permission of AIP Publishing.}
    \label{fig:speedup_a_vs_ab}
\end{figure}

In the context of coupled-cluster (CC) theory,\cite{Shavitt09} CD reduces the prefactor of the computational cost but not the overall scaling. Its most significant advantage, however, is the substantial reduction in both integral storage and memory bandwidth requirements. In \cfour, we have implemented a CD-based CCSD\cite{Purvis82} algorithm that fully exploits Abelian point-group symmetry.\cite{Nottoli22b} The working equations are formulated in a spin-adapted form, with the definition of the intermediate quantities largely following the work of Stanton {\it et al.}\cite{Stanton91a} For computational efficiency, all integral classes involving one or two virtual indices (e.g., $\langle Ab|Ij\rangle$) are reconstructed on the fly from the CVs, where uppercase letters denote $\alpha$ molecular orbitals and lowercase letters denote $\beta$ molecular orbitals. The rate-determining step in a conventional CCSD calculation is given by the so-called particle--particle ladder (PPL) contribution
\begin{equation}\label{eq:PPL}
    Z_{Ij}^{Ab} = \sum_{Ef}\tau_{Ij}^{Ef}\mathscr{W}_{AbEf},
\end{equation}
where $\tau_{Ij}^{Ab} = t_{Ij}^{Ab} + t_{I}^At_J^B$ denotes the usual combination of double- and single-excitation amplitudes. The $\mathscr{W}$ intermediate can be rewritten in terms of CVs and reads 
\begin{equation}
    \mathscr{W}_{AbEf} = \sum_P\left[\left(L^P_{AE}-t_{AE}^P\right)L^P_{bf} - \sum_m t_m^bL^P_{AE}L^P_{mf}\right],
\end{equation}
with $t_{AE}^P = \sum_M t_M^AL^P_{EM}$. The formal computational scaling of the PPL term is $\mathcal{O}(O^2V^4)$. To improve upon this, the contraction is performed using the symmetric/antisymmetric algorithm~\cite{Saebo87} that is
\begin{equation}
     Z_{Ij}^{Ab} = S_{Ij}^{Ab} + A_{Ij}^{Ab},
\end{equation}
where
\begin{align}
    &S_{Ij}^{Ab} = \sum_{Ef}{}^+\tau_{Ij}^{Ef}{}^+\mathscr{W}_{AbEf}, \\
    &A_{Ij}^{Ab} = \sum_{Ef}{}^-\tau_{Ij}^{Ef}{}^-\mathscr{W}_{AbEf}, \\
    &{}^\pm\tau_{Ij}^{Ef} = \frac{1}{2}\left(\tau_{Ij}^{Ef}\pm\tau_{Ij}^{Fe}\right), \\
    &{}^\pm\mathscr{W}_{AbEf} = \frac{1}{2}\left(\mathscr{W}_{AbEf}\pm\mathscr{W}_{AbFe}\right).
\end{align}
This reduces the computational prefactor by a factor of four.
The implementation avoids storing quantities that scale as $V^4$ or $V^3O$ (with $V$ denoting the number of virtual orbitals) by reconstructing only the batches of integrals required for each contraction. For the PPL contribution, this is achieved by fixing the index $A$ (called ``loop $a$'' strategy) in Eq.~\eqref{eq:PPL} and distributing the computation over OpenMP threads, such that only $V^3N_{\rm threads}$ quantities need to be reconstructed at any given time. All contractions are carried out using optimized \texttt{DGEMM} routines to maximize computational efficiency. In addition, the implementation provides a lower-memory algorithm in which both indices $A$ and $B$ are fixed, reducing the memory requirement to $V^2N_{\rm threads}$ (``loop $ab$'' strategy). The effectiveness of the parallelization for both algorithms is illustrated in Fig.~\ref{fig:speedup_a_vs_ab}, using coronene as a test case with the cc-pVTZ basis set\cite{Dunning89} and Abelian point-group symmetry enforced. The calculations were carried out on a single Intel Xeon Gold 6140M node equipped with 1.1 TB of RAM. Both algorithms exhibit similar speedup behavior with ``loop $a$'' performing slightly better at low thread count but reaching a plateau earlier. ``Loop $ab$'' scales somewhat better due to its smaller memory footprint, which improves cache utilization. 
Overall, both approaches achieve speedups of approximately 8-10 using 16 threads, after which their performance plateaus.
Finally, the efficiency of the implementation was demonstrated by performing a calculation on buckminsterfullerene (C$_{60}$) using the cc-pVTZ basis set,\cite{Dunning89} which consists of 1800 basis functions (1620 virtual and 180 doubly occupied orbitals). The frozen-core approximation was employed together with a CD threshold of $10^{-4}$. Notably, the complete set of CVs required only 29~GB of memory, compared with approximately 1~TB that would be required to store the unique elements of the $\langle Ab|Ef\rangle$ integrals explicitly. Using the ``loop a'' algorithm, a single iteration took about 2~h, and the calculation converged in 1~day 19~h to 10$^{-7}$ in the maximum norm of the residual, on the same Intel Xeon Gold node mentioned above.

\section{Computation of molecular structures and properties using Cholesky decomposition}
\subsection{Analytical gradients with Cholesky decomposition}
\label{section6}
As the derivatives of the ERIs 
do not constitute a positive semidefinite matrix, they cannot be directly decomposed via a CD. However, it is possible to derive CD-type expressions for them via differentiation of the CD expressions for the undifferentiated ERIs\cite{Feng19, Burger21, Gauss23, Burger25} or, alternatively, by exploiting the formal equivalence of RI/DF and CD.\cite{Aquilante08c, Bostroem14, Delcey14}

The former strategy requires the explicit computation of differentiated CVs. Unlike in the case of magnetic perturbations,\cite{Burger21} the handling of perturbed CVs is cumbersome when dealing with geometrical derivatives, as the CVs depend on all perturbations. The implementation of CD-based molecular gradients within {\cfour} therefore avoids their construction and instead is based on the exploitation of the formal equivalence between the RI/DF and the CD approximations of the ERI matrix.\cite{Pedersen09}

The expression for the geometrical gradient for a general wavefunction-based method takes the form:
\begin{equation}
    \label{eq:2el_gradient}
    \frac{dE}{dx} = \sum_{\mu \nu} D_{\mu \nu} h_{\mu \nu}^x + \frac{1}{2}\sum_{\mu \nu \rho \sigma} \Gamma_{\mu \rho \nu \sigma} (\mu \nu \vert \rho \sigma)^x + \sum_{\mu \nu} I_{\mu \nu} S_{\mu \nu}^x,
\end{equation}
where $D_{\mu \nu}$ is the one-body density matrix, $\Gamma_{\mu \rho \nu \sigma}$ is the two-body density matrix, $I_{\mu \nu}$ is the Lagrange multiplier associated with the enforcement of the orthonormality between MOs, $h_{\mu \nu}^x$ is the explicit derivative of the one-electron Hamiltonian, $(\mu \nu \vert \rho \sigma)^x$ is the derivative of the ERI matrix, and $S_{\mu \nu}^x$ is the derivative of the overlap matrix.
In this work, we will focus solely on the two-electron part of the analytical gradient, as it is the only one affected by the use of CD.
By following the strategy detailed above and taking the derivative of the RI/DF expression for the ERI matrix (i.e., Eq.~\ref{eq:DF1}), we can reformulate the second term in Eq.~\ref{eq:2el_gradient} as follows:
\begin{equation}
    \label{eq:twoelderint}
    \frac{1}{2}\sum_{\mu \nu \rho \sigma} \Gamma_{\mu \rho \nu \sigma} (\mu \nu \vert \rho \sigma)^x = \sum_{\mu \nu} \sum_P (\mu \nu \vert P)^x J_{\mu \nu}^P + \sum_{\rho \sigma} \sum_P \Bar{J}_{\rho \sigma}^P (P \vert \rho \sigma)^x - \sum_{PQ} W_{PQ} (P \vert Q)^x,
\end{equation}
where $J_{\mu \nu}^P$, $\Bar{J}_{\rho \sigma}^P$ and $W_{PQ}$ are effective densites. Their definitions differ depending on the method of choice. 
The effective densities are contracted on-the-fly with differentiated two-electron integrals: if either product density $\vert \munu )$ or $\vert \rho \sigma )$ belongs to the Cholesky basis, it follows that the considered perturbed integral is the element $(\munu \vert P)^x$ (or $(\rho \sigma \vert P)^x$) of the differentiated non-orthogonal CVs and is thus contracted with $J_{\mu \nu}^P$ and $\Bar{J}_{\munu}^P$ (or $J_{\rho \sigma}^P$ and $\Bar{J}_{\rho \sigma}^P$); otherwise, if both product densities correspond to Cholesky indices, the integral is the element $(P \vert Q)^x$ of the perturbed Cholesky metric and contracted with $W_{PQ}$.

In the {\cfour} program package, a CD-based implementation of analytical geometrical gradients is available at the SCF, CASSCF, MP2, and CCSD\cite{Melega26} levels of theory, and is based on the {\mint} integral package.\cite{mint} In the case of SCF, the two-body density matrix is separable and can be written in terms of products of the one-body density matrix and therefore poses no particular issue for an effective implementation. For CASSCF, in constrast, the \textit{all active} block of the two-body density matrix (denoted with indices $u, v, x, y$ in the following) is non-separable. To avoid a full transformation of the density to the AO basis, we define the following intermediates in the MO basis, where in particular only active indices are considered
\begin{align}
    &\tilde{L}^P_{xy} = \sum_{\rho\sigma}C_{\rho x}C_{\sigma y}\tilde{L}^P_{\rho\sigma}, \label{eq:lt_act_cas}\\
    &\Gamma^P_{uv} = \frac{1}{2} \sum_{xy}\Gamma_{uvxy}\tilde{L}^P_{xy},\label{eq:dt_act}
\end{align}
where $C$ are the MO expansion coefficients and we have defined $\textit{transformed}$ CVs $\tilde{L}_{\munu}^P$:
\begin{equation}
    \tilde{L}_{\mu \nu}^P = \sum_Q L_{\mu \nu}^Q M_{QP}^{-1},
\end{equation}
with $M_{PQ}$ being the Cholesky factor of the Cholesky-basis metric.
Intermediates \eqref{eq:lt_act_cas} and \eqref{eq:dt_act} are then used to build effective densities
\begin{align}
    &J^P_{\mu\nu} = \sum_{uv}C_{\mu u} C_{\nu v}\Gamma^P_{uv} = \bar{J}^P_{\mu \nu}, \\
    &W_{PQ} = \sum_{uv} \tilde{L}^P_{uv} \Gamma^P_{uv}.
\end{align}


For CD-CCSD gradient calculations, {\cfour} computes the effective density intermediates according to the following definitions:
\begin{align}
    &\tilde{L}_{qs}^P = \sum_{\rho \sigma} C_{\rho q} C_{\sigma s} \tilde{L}_{\rho \sigma}^P,\\
    &\Gamma_{pr}^P = \sum_{qs} \widetilde{\Gamma}_{pqrs}\tilde{L}_{qs}^P, \label{eq:d2xtl1}\\
    &\bar{\Gamma}_{qs}^P = \sum_{pr} \widetilde{\Gamma}_{pqrs}\tilde{L}_{pr}^P, \label{eq:d2xtl2}\\
    &J_{\mu \nu}^P = \sum_{pr} C_{\mu p} C_{\nu r} \Gamma_{pr}^P,\\
    &\bar{J}_{\rho \sigma}^P = \sum_{qs} C_{\rho q} C_{\sigma s} \bar{\Gamma}_{qs}^P,\\
    &W_{PQ} = \sum_{\mu \nu} \tilde{L}_{\mu \nu}^P J_{\mu \nu}^Q.
\end{align}

Since {\cfour} currently only offers CD-CCSD gradients based on an RHF reference, the two-body density matrix is  available in its spin-adapted form $\widetilde{\Gamma}_{pqrs} = 2\Gamma_{PqRs} - \Gamma_{PqSr}$ and the factor $\tfrac{1}{2}$ is absorbed into the definiition of the spin-adapted density. The main benefit of using CD in this case is that it allows us to avoid the full transformation of the two-body density from the MO to the AO basis. In fact, each block of the density is contracted separately. Moreover, within this formalism, the contractions involving the $vvvv$ and $vvvo$ blocks of $\widetilde{\Gamma}_{pqrs}$ are performed on-the-fly, to avoid storing in memory $V^4$- and $V^3O$-scaling arrays.

\begin{figure}
    \centering
    \includegraphics[width=0.25\linewidth]{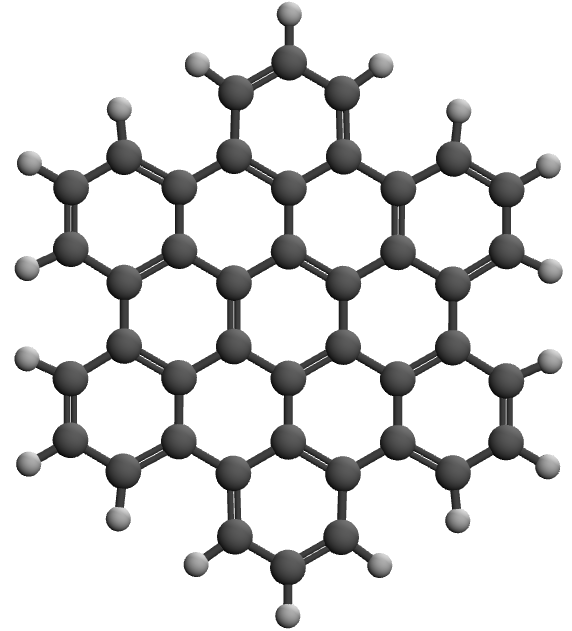}
    \caption{Structure of the hexabenzocoronene molecule, optimized at the CD-CCSD level.}
    \label{fig:hbc_geometry}
\end{figure}

\begin{figure}[h]
\centering
\includegraphics[scale=0.6,angle=0]{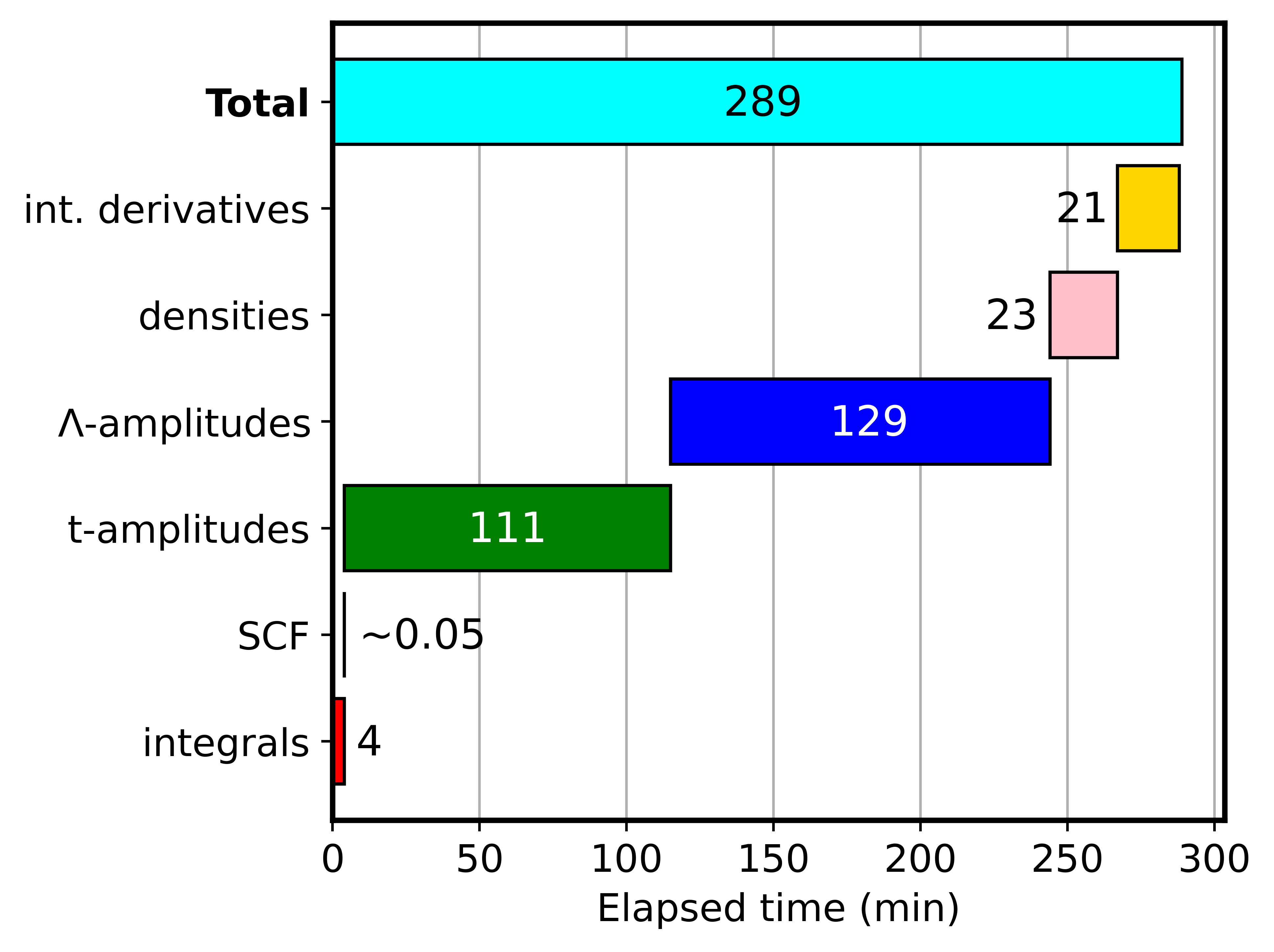}
\caption{Gantt chart for the timings (expressed in minutes) of each step required for the computation of CD-CCSD gradients for hexabenzocoronene in the cc-pVDZ basis. Reproduced from Ref. \citenum{Melega26}. Published by the American Chemical Society under a CC BY 4.0 license.}
\label{fig:gantt_hbc}
\end{figure}

The CD-CCSD gradient implementation in {\cfour} has been used to perform the geometry optimization of the hexabenzocoronene molecule (shown in Fig.~\ref{fig:hbc_geometry}), on a node equipped with two AMD EPYC 7282 16 Core processor (2.8 GHz) and 512 GB of RAM and requesting 32 OpenMP threads. The calculation was carried out using Dunning's cc-pVDZ basis set,\cite{Dunning89} which consists of 678 basis functions, 135 occupied MOs, and 543 virtual MOs. The calculation exploited \(D_{2h}\) symmetry, the largest Abelian subgroup of the \(D_{6h}\) point group of the initial geometry. The two-step CD algorithm yielded 3710 CVs on average during the optimization (with a CD threshold equal to $10^{-4}$), almost evenly distributed among the 8 irreducible representations. The optimization procedure converged to the equilibrium geometry in 24 steps, taking about 4 days and 15 hours in total. The timings associated with the individual tasks in the computation of gradients in the first optimization step are shown in the Gantt chart in Fig.~\ref{fig:gantt_hbc}. The peak memory consumption associated with a single optimization step was 302 GB.

\begin{figure}
    \centering
    \includegraphics[scale=0.6]{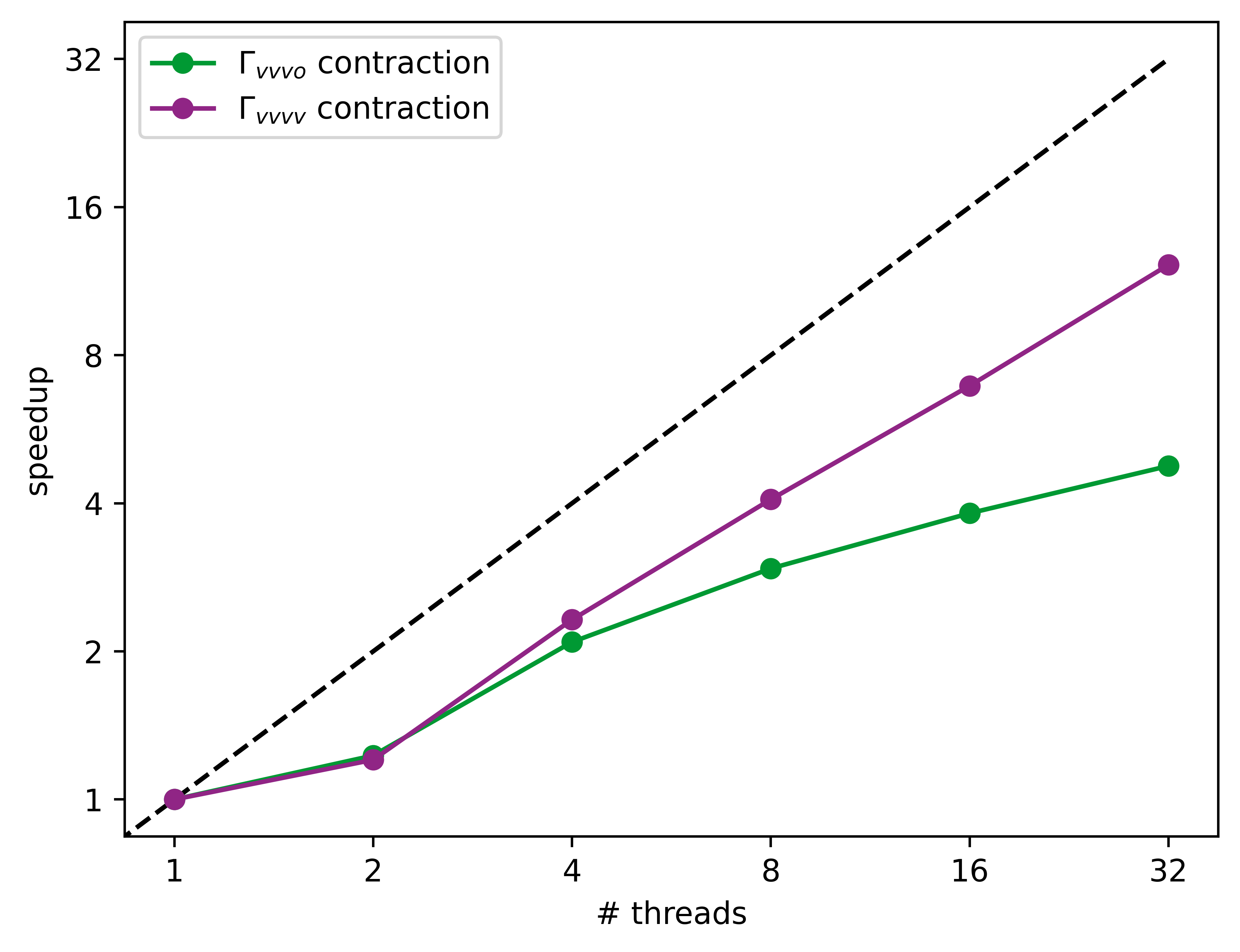}
    \caption{Speedup graph for CD-CCSD/cc-pVDZ gradient calculations on coronene. The green curve refers to the contraction involving the \textit{vvvo} block of the two-body density matrix, while the purple curve refers to the contraction involving the \textit{vvvv} block of the two-body density matrix, both in the evaluation of the molecular gradient. The ratio between the execution time of the serial code and the execution time of the parallel one is plotted against the number of requested OpenMP threads, in the log$_2$ scale. Reproduced from Ref. \citenum{Melega26}. Published by the American Chemical Society under a CC BY 4.0 license.}
    \label{fig:speedup_coronene}
\end{figure}

We also assessed the parallel performance of the CD-CCSD gradient implementation. We have run several calculations on the coronene molecule using the cc-pVDZ basis~\cite{Dunning89} on a compute node equipped with two AMD EPYC 7282 16 Core processor (2.8 GHz) and 256 GB of RAM, each employing a different number of shared-memory OpenMP threads (in particular, 1, 2, 4, 8, 16, and 32). Fig.~\ref{fig:speedup_coronene} reports the speedup, defined as the ratio of the serial to parallel wall time. This analysis has been carried out on the contractions between the two-electron integral derivatives and the \textit{vvvo} (reported in the green curve) and the \textit{vvvv} (reported in the purple curve) blocks of the CCSD two-body density matrix, which feature respectively $\mathcal{O}(O^3V^3)$- and $\mathcal{O}(O^2V^4)$-scaling terms. Both contractions are performed within a parallelized outermost loop over virtual MO indices. The speedup plot related to the former deviates from the ideal behavior approaching a plateau at 16 OpenMP threads, likely due to the complexity of the implementation of the $\widetilde{\Gamma}_{abci}$ contraction, which contains significant serial sections. 
For the vvvv contraction, the speedup continues to increase almost linearly up to 32 threads, although it already deviates from the ideal speedup at low thread counts, likely due to the remaining serial sections of the code. No indication of a plateau is observed at 32 threads.

CD itself does not affect the computational scaling of CCSD gradient calculations. Despite this, it has a notable impact on the overall performance for two reasons. First, the cost of integral transformations becomes negligible, whereas it acts as a relevant bottleneck in traditional implementations due to disk I/O operations and repeated integral transformations. Second, it greatly reduces the RAM requirements in the CC step of the calculations, thus further improving their scalability. To showcase this aspect, we performed a single-point energy and gradient calculation for anthracene using the CD-based code and two variants of the traditional CFOUR implementation: the standard MO algorithm, which requires storage of the full vvvv block of two-electron integrals for optimal efficiency, and the partial AO-direct algorithm, which represents a more realistic choice when memory is limited.
The calculations were performed on a computer node equipped with two AMD EPYC 7282 16 Core processors (2.8 GHz) and 512 GB of RAM.
\begin{table}[H]
    \centering
    \begin{tabular}{lrrr}
    \toprule
    Step & CD & Traditional - MO & Traditional - AO\\
    \midrule
    Integrals & 199 & 447 & 447 \\
    SCF & 2 & 264 & 264 \\
    AO2MO & - & 931 & 188 \\
    CC & 973 & 1994 & 4237 \\
    $\Lambda$ & 1245 & 1490 & 3865\\
    Gradient & 1068 & 2783 & 4808 \\
    Total & 3388 & 8473 & 13824 \\
    \bottomrule
    \end{tabular}
    \caption{Timings (measured in seconds) for a CCSD energy and gradient calculation for anthracene comparing the CD implementation and the traditional implementation using both a full MO and a partial AO algorithm. Integrals: cost of the evaluation of the molecular integrals or their Cholesky decomposition; SCF: cost of solving the SCF problem using a second-order solver; AO2MO: cost of the integral transformations (for the CD code, the cost is included in the timings reported for the $t$-amplitudes); CC: cost to solve the CCSD equations; $\Lambda$: cost to solve the Lambda equations; Gradient: cost to assemble the one- and two-body reduced density matrices and to contract them with the integral derivatives.}
    \label{tab:cfr}
\end{table}
As shown in Table~\ref{tab:cfr}, all steps of the calculation are less expensive when exploiting CD, including the CC steps, despite their unchanged formal scaling. Overall, the CD calculation is approximately 2.5 and 4 times faster than the conventional MO and partial AO-direct calculations, respectively.

\subsection{Computation of magnetic properties with Cholesky decomposition\label{sec:magnetic}}
\label{section7}
\cfour\ also offers the opportunity to compute magnetic properties such as nuclear magnetic resonance (NMR) shielding and magnetizability tensors\cite{Gauss95c,Gauss02c,Helgaker12} using CD. 
Both quantities can be obtained as the second derivative of the energy with respect to an external magnetic field $\bf B$ and/or a nuclear magnetic moment ${\bf m}^K$, or as the second derivative of the energy with respect to the external magnetic field $\bf B$, respectively.
\begin{equation} \label{eq:Shieldingtensor}
    \sigma_{ij}^K = \left( \frac{\mathrm{d}^2E}{\mathrm{d}m^K_i \mathrm{d} B_j} \right)_{\mathbf{B},\mathbf{m}^K=0},
\end{equation}
\begin{equation} \label{eq:Magtensor}
    \xi_{ij} = -\left( \frac{\mathrm{d}^2E}{\mathrm{d}B_i \mathrm{d} B_j} \right)_{\mathbf{B}=0}, 
\end{equation}
with $i$ and $j$ referring to $x$, $y$, or $z$. The expressions resulting from Eqs.~\ref{eq:Shieldingtensor} and \ref{eq:Magtensor} involve derivatives of the AO integrals with respect to $\bf B$ and ${\bf m}^K$. Our focus in the following is on the derivatives of the ERIs that appear when using gauge-including atomic orbitals (GIAOs, also known as London orbitals).\cite{London37,Hameka58,Ditchfield72,Helgaker91,Wolinski90,Gauss92} The latter are given by
\begin{eqnarray}
\label{giao}
\omega_{\mu}({\bf r},{\bf B}) = \exp\left( - \frac{i}{2} ({\bf B} \times({\bf R}_{\mu} - {\bf R}_O) \cdot {\bf r} \right)\chi_{\mu}({\bf r})
\end{eqnarray}
with $\chi_{\mu}({\bf r})$ as the usual field-independent AOs, ${\bf R}_{\mu}$ as the center of the basis function, ${\bf R}_O$ as the chosen gauge origin, and $\bf B$ as the external magnetic field. GIAOs are used to ensure gauge-origin independence for the computed magnetic properties\cite{Pulay93} for the price of introducing a dependence of the AOs on the magnetic field. The latter is the reason why the calculation of magnetic properties involve first derivatives (and second derivatives in the case of magnetizabilities) of the ERIs with respect to the components of the magnetic field. 
Unlike the unperturbed ERIs, these derivative integrals do not represent a positive semidefinite matrix and thus cannot be decomposed using a CD. However, it is possible to obtain a decomposition for these integrals (in the following referred to as NMR and magnetizability integrals) by differentiating the equations for the unperturbed ERIs with respect to the magnetic field components.
\cite{Burger21,Burger25} Hence, the matrix for the first derivatives (NMR integrals) can be decomposed via\cite{Burger21,Gauss23} 
\begin{equation} \label{CD-NMR-Int}
    \frac{\partial (\sigma \rho | \mu \nu)}{\partial B_i} \approx \sum^{M}_{P} \left(\frac{\partial L_{\sigma \rho}^{P}}{\partial B_i}  L_{\nu \mu}^{P} - L_{\sigma \rho}^P  \frac{\partial L_{\nu \mu}^{P}}{\partial B_i}\right).
\end{equation}
To derive the expression for the perturbed CVs,
\begin{equation} \label{eq:NMRCV}
    \dfrac{\partial L_{\sigma \rho}^P}{\partial B_i}=(\widetilde{\nu \mu| \mu \nu})^{-\frac{1}{2}} \left\{ \left(\frac{\partial \siro}{\partial B_i} \Bigg| \mu \nu \right) -\sum_{Q=1}^{P-1} \dfrac{\partial L_{\sigma \rho}^Q}{\partial B_i} L_{\nu \mu}^{Q} \right\},
\end{equation} 
the equation for the unperturbed CVs is differentiated. Since the NMR integrals and the corresponding perturbed CVs are purely imaginary, the derivatives of the updated diagonal elements vanish and do not need to be considered in Eq. \ref{eq:NMRCV}. 
In addition, as the NMR integrals have only fourfold permutational symmetry (instead of eightfold permutational symmetry as the unperturbed ERIs) and in particular lack symmetry with respect to an interchange of the two AOs on the bra and ket side, it turns out advantageous to reintroduce some of permutational symmetry. For this reason, the NMR 
integrals are split into their partial derivatives\cite{Kollwitz96,Gauss23},
\begin{equation} \label{SplitNMRInt}
    \frac{\partial (\siro|\munu)}{\partial B_i}=\left(\frac{\partial \siro}{\partial B_i} \Bigg| \munu \right) + \left( \siro \Bigg| \frac{\partial \munu}{\partial B_i} \right),
\end{equation}
with the following permutation relations now holding
\begin{equation} \label{eqn:SNI2}
    \left(\frac{\partial \siro}{\partial B_i} \Bigg| \munu \right) = \left(\frac{\partial \siro}{\partial B_i} \Bigg| \numu \right), \quad
    \left( \siro \Bigg| \frac{\partial \munu}{\partial B_i} \right) = - \left( \siro \Bigg| \frac{\partial \numu}{\partial B_i} \right).
\end{equation}
As already shown in Refs.~\citenum{Burger21} and \onlinecite{Gauss23}, the perturbed CVs are antisymmetric,
\begin{equation} \label{SignChangeNMR}
    \frac{\partial L_{\sigma \rho}^{P}}{\partial B_i} = -\frac{\partial L_{\rho \sigma}^{P}}{\partial B_i}.
\end{equation}
It is important to realize that this procedure to obtain a decomposition of the NMR integrals does not constitute an independent CD. Actually, the NMR integrals are represented in the Cholesky basis constituted by the unperturbed ERIs, which implies that the CVs in Eq.~\ref{eq:NMRCV} are assigned to an unperturbed $\munu$-pair. Within the two-step algorithm,\cite{Folkestad19,Zhang21} this means that the first step (i.e., the calculation of the Cholesky basis) remains the same, while the perturbed CVs are obtained via
\begin{equation} \label{TwoStep-NMRCV}
    \frac{\partial L_{\siro}^P}{\partial B_i} = \sum_Q \left(\frac{\partial \siro}{\partial B_i} \bigg| Q \right) (Q|P)^{-\frac{1}{2}}
\end{equation}
in the second step. \\
In the case of the magnetizability integrals, the procedure for deriving the equations is the same, that is straightforward differentiation of the expression for the unperturbed ERIs. The magnetizability integrals are then given by
\begin{equation} \label{CD_Mag_matrix}
    \dfrac{\partial^2 (\sigma \rho |\mu \nu )}{\partial B_i \partial B_j} \approx \sum^{M}_{P} \left( \dfrac{\partial^2 L_{\sigma \rho}^{P}}{\partial B_i \partial B_j}  L_{\nu \mu}^{P} - \dfrac{\partial L_{\sigma \rho}^{P}}{\partial B_i}  \dfrac{\partial L_{\nu \mu}^{P}}{ \partial B_j} 
- \dfrac{\partial L_{\sigma \rho}^{k}}{\partial B_j} \dfrac{\partial L_{\nu \mu}^{P}}{ \partial B_i} + L_{\sigma \rho}^P  \dfrac{\partial^2 L_{\nu \mu}^{P}}{\partial B_i \partial B_j} \right), 
\end{equation}
where the doubly perturbed CVs are calculated via 
\begin{equation} \label{eq:MagCV_short}
    \dfrac{\partial^2 L_{\sigma \rho}^P}{\partial B_i \partial B_j} =(\widetilde{\nu \mu| \mu \nu})^{-\frac{1}{2}} \left\{  \left(\dfrac{\partial^2 \sigma \rho}{\partial B_i \partial B_j} \bigg| \mu \nu \right)  
    -\sum^{P-1}_{Q=1}  \dfrac{\partial^2 L_{\sigma \rho}^{Q}}{\partial B_i \partial B_j} L_{\nu \mu}^{Q}  \right\}.
\end{equation}
Here, the perturbation dependence of the Cholesky basis functions is neglected, which leads to an error in the reconstructed integrals,\cite{Burger25} but this error has been shown to be very small 
and, thus, the reconstructed integrals are of sufficient accuracy for the computation of the magnetizability tensor.\cite{Burger25} Unlike the NMR integrals and the perturbed CVs, the magnetizability integrals and double perturbed CVs are real and by splitting the magnetizability integals,
\begin{equation} 
    \dfrac{\partial^2 (\sigma \rho| \mu \nu)}{\partial B_i \partial B_j} = \left(\dfrac{\partial^2 \sigma \rho}{\partial B_i \partial B_j} \Bigg| \mu \nu \right) + \left( \dfrac{\partial \sigma \rho}{\partial B_i} \Bigg| \dfrac{\partial \mu \nu}{\partial B_j} \right) + \left( \dfrac{\partial \sigma \rho}{\partial B_j} \Bigg| \dfrac{\partial \mu \nu}{\partial B_i} \right) + \left( \sigma \rho \Bigg| \dfrac{\partial^2 \mu \nu}{\partial B_i \partial B_j} \right),
\label{mag1}
\end{equation}
permutational symmetry again can be exploited.
The second and third terms in Eq.~(\ref{mag1})
are antisymmetric with respect to an interchange of the AO indices for electron 1 and 2, while the first and fourth term are symmetric,
\begin{equation} \label{eqn:PSM2} 
    \left(\frac{\partial^2 \siro}{\partial B_i \partial B_j} \Bigg| \munu \right) = \left(\frac{\partial^2 \siro}{\partial B_i \partial B_j} \Bigg| \numu \right), \quad
    \left( \siro \Bigg| \frac{\partial^2 \munu}{\partial B_i \partial B_j} \right) = \left( \siro \Bigg| \frac{\partial^2 \numu}{\partial B_i \partial B_j} \right). 
\end{equation}
It follows that the doubly perturbed CVs are symmetric with respect to an interchange of the indices, 
\begin{equation}
    \dfrac{\partial^2 L_{\sigma \rho}^P}{\partial B_i \partial B_j}=\dfrac{\partial^2 L_{\rho \sigma}^{P}}{\partial B_i \partial B_j}.
\end{equation}

In case of the NMR integrals, the two-step algorithm is used by default.\footnote[1]{For historical reasons, the one-step algorithm is also available in \cfour.}
NMR shielding calculations using CD can be performed at the HF, MP2\cite{Burger21}, and at the CASSCF level.\cite{Nottoli22b} 
Current work focuses on an extension of the capability in \cfour\ to compute NMR shieldings for CC methods (CC shieldings without CD have been reported in Refs.
\onlinecite{Gauss95a}, \onlinecite{Gauss95b}, and \onlinecite{Gauss96a}).
The magnetizability tensor so far can be computed only at the HF level using the one-step CD procedure.\cite{Burger25} Future work will focus on the implementation of the two-step algorithm for the magnetizability integrals and on CD based calculation of magnetizabilities at MP2, CC, and CASSCF level. 

The required NMR and magnetizability integrals for their CD are computed using the McMurchie-Davidson scheme\cite{McMurchie78} as implemented in the \mint\ integral package.\cite{mint} During the CD of the integrals, 
the whole set of unperturbed and (one set of) perturbed CVs are kept in memory. To compute the NMR and magnetizability tensor, all perturbed and unperturbed two-electron integrals appearing in the corresponding equations (see, for example, Ref.~\onlinecite{Burger21}) are replaced by their CD counterparts. However, unperturbed and perturbed integrals with less than three virtual indices are explicitly formed and stored, while all other integrals are represented using the corresponding Cholesky vectors. Abelian point-group symmetry can be exploited in the computation of CD based NMR shieldings.


\begin{figure} [H]
    \centering
    \includegraphics[width=0.55\linewidth]{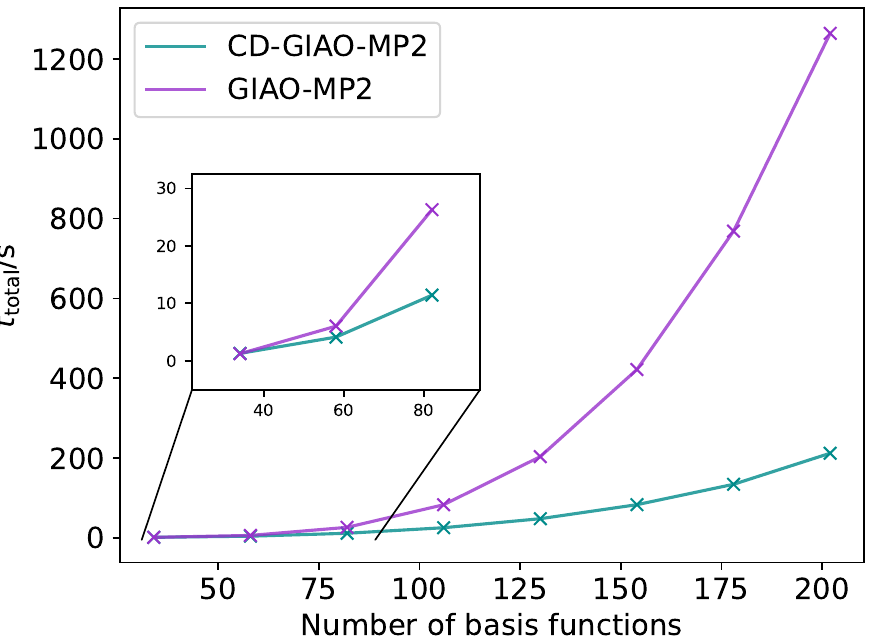}
    \caption{Total timings $t_{\mathrm{total}}$ for NMR shielding calculations of the homologous series of straight-chained alkanes $\mathrm{C}_2\mathrm{H}_{2\mathrm{n}+2}$ in a cc-pVDZ basis\cite{Dunning89} using a Cholesky threshold $\tau$ of $10^{-5}$, with and without exploitation of CD.} 
    \label{fig:BreakEvenPoint}
\end{figure}
\begin{figure} [H]
    \centering
    \includegraphics[width=11cm]{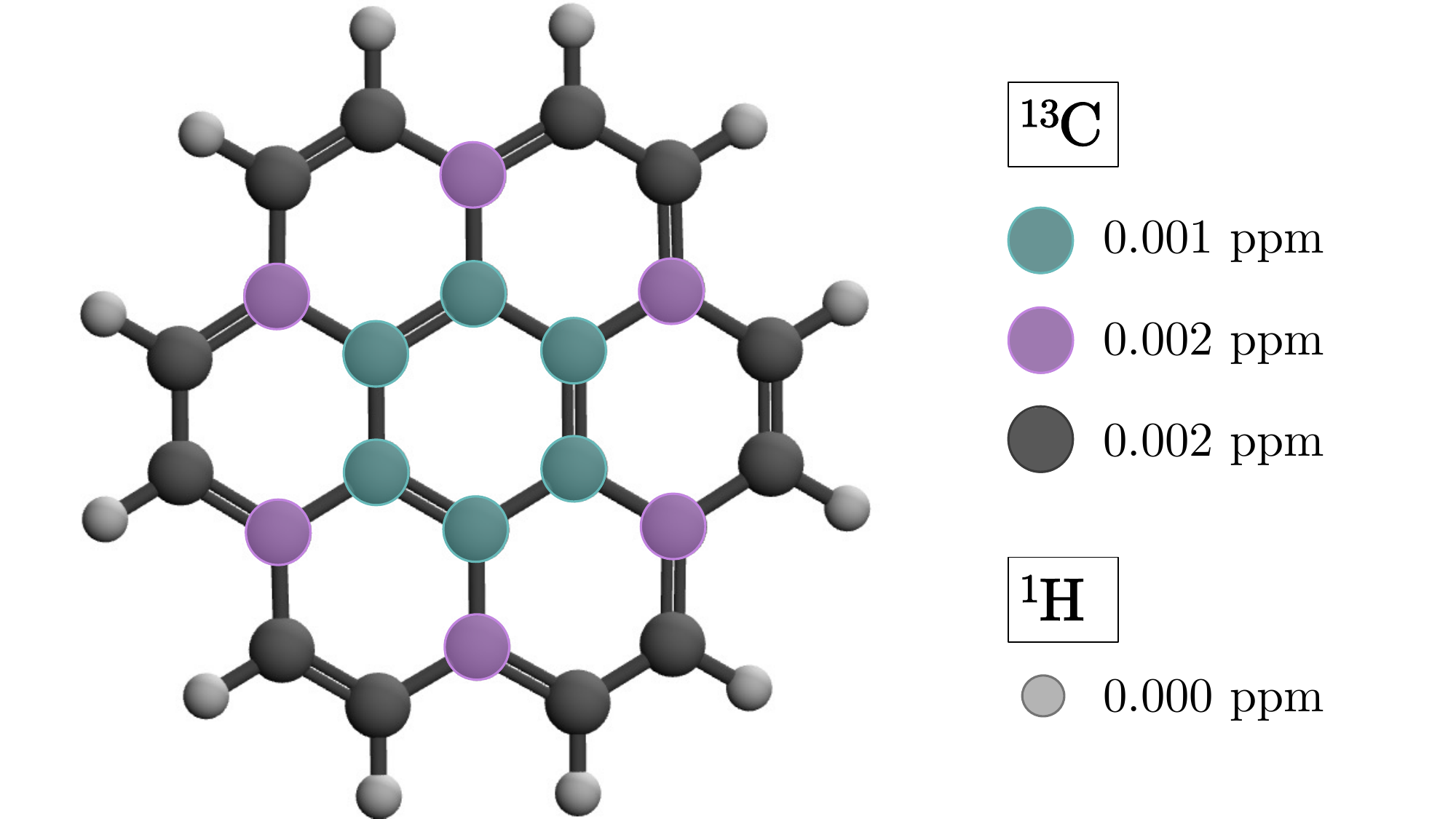}
    \caption{Maximum errors (given in ppm) in the $^{13}$C and $^1$H NMR shieldings (with respect to a standard GIAO-MP2 calculation) for coronene in CD-GIAO-MP2 computations using a Cholesky threshold $\tau$ of $10^{-5}$ and the dzp basis set.}
    \label{CoroneneNMRerror}
\end{figure}

Calculations on the homologous series of straight-chained alkanes $\mathrm{C}_2\mathrm{H}_{2\mathrm{n}+2}$ were performed to investigate the break-even point between the standard GIAO-MP2 and the CD-based schemes. The total computational timings are presented in Fig.~\ref{fig:BreakEvenPoint}. For the smallest system, i.e., methane with 34 basis functions, both implementations show comparable computational costs. For the next-largest system (ethane with 58 basis functions), however, CD already provides a reduction in computational time and, as expected, the computational savings increase with system size. The results in Fig. \ref{fig:BreakEvenPoint} indicate that CD becomes advantageous for systems with approximately 50 basis functions or more.
The accuracy of the NMR shielding constants computed at the CD-GIAO-MP2 level is determined by comparing them to standard GIAO-MP2 computations. This comparison is performed for coronene with a dzp basis set, and the results are presented in Fig. \ref{CoroneneNMRerror}. The maximum errors shown in Fig. \ref{CoroneneNMRerror} demonstrate that the approximations introduced by CD result in only minor deviations. A Cholesky threshold $\tau$ of $10^{-5}$ provides an accuracy that is sufficient for chemical applications. 

\begin{figure}[H]
    \centering
    \begin{subfigure}{0.3\textwidth}
        \centering
        \includegraphics[width=0.7\textwidth]{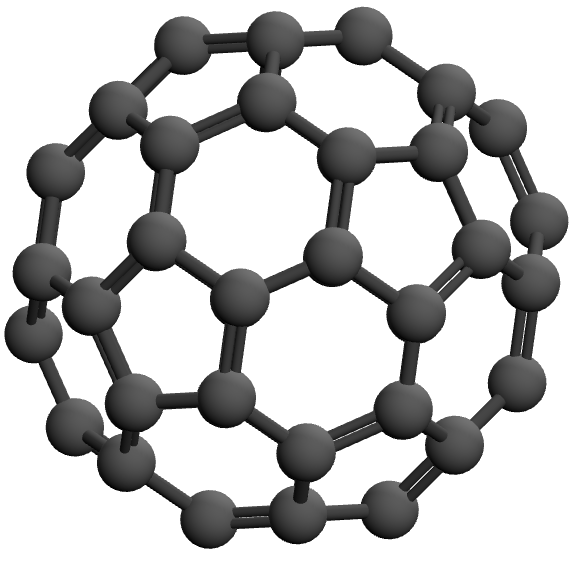}
        \caption{C$_{60}$ \\  $\mathrm{I_{h}}$ ($\mathrm{D_{2h}}$)}
    \end{subfigure}
    \hfill
    \begin{subfigure}{0.3\textwidth}
        \centering
        \includegraphics[width=0.7\textwidth]{HBC.png}
        \caption{HBC \\ $\mathrm{D_{6h}}$ ($\mathrm{D_{2h}}$)}
    \end{subfigure}
    \hfill
    \begin{subfigure}{0.3\textwidth}
        \centering
        \includegraphics[width=0.7\textwidth]{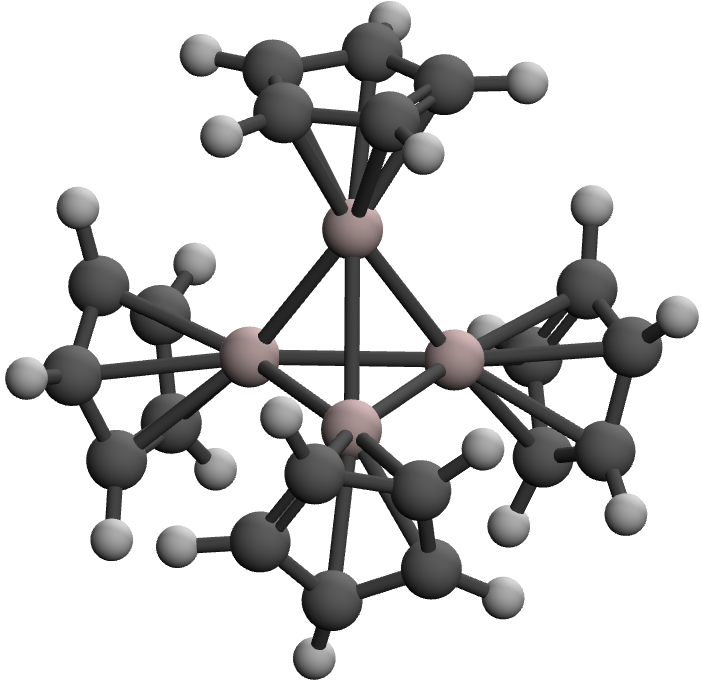}
        \caption{Al$_4$Cp$_4$ \\ $\mathrm{D_{2d}}$ ($\mathrm{D_{2}}$)}
    \end{subfigure}
    \vspace{0.5cm}
    \begin{subfigure}{0.3\textwidth}
        \centering
        \includegraphics[width=0.7\textwidth]{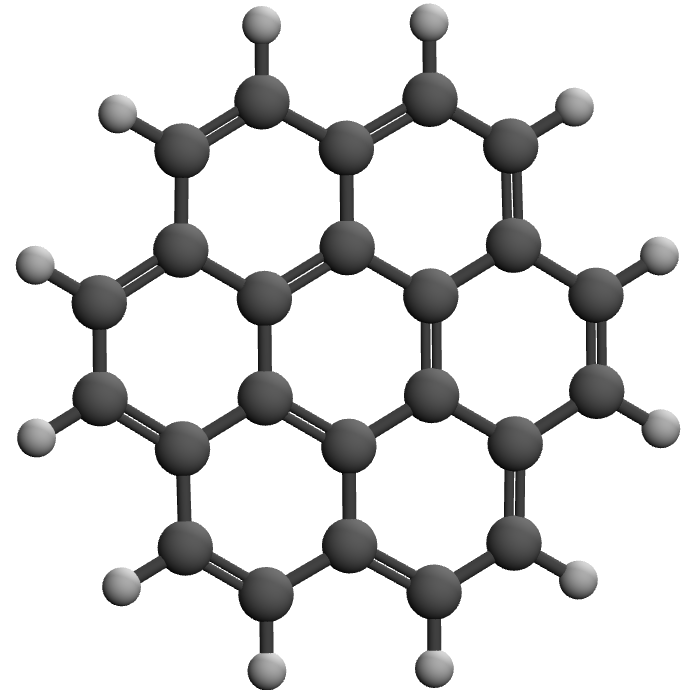}
        \caption{Coronene \\  $\mathrm{D_{6h}}$ ($\mathrm{D_{2h}}$)}
    \end{subfigure}
    \hfill
    \begin{subfigure}{0.3\textwidth}
        \centering
        \includegraphics[width=0.7\textwidth]{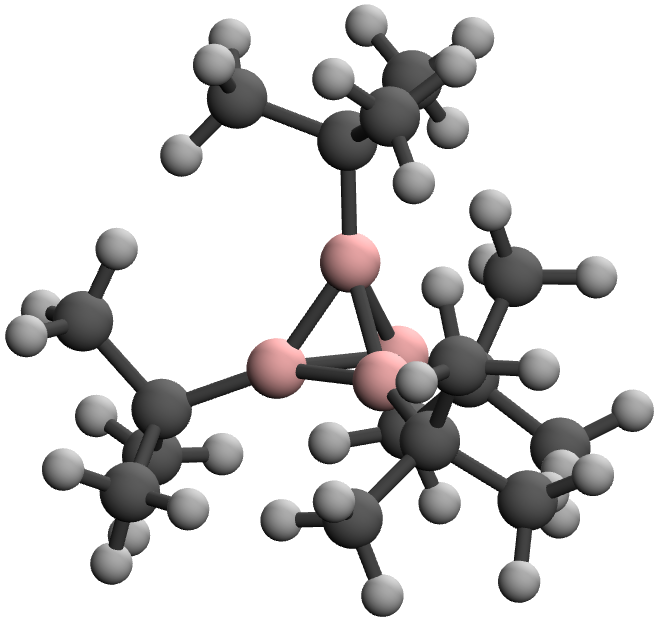}
        \caption{B$_4$tBu$_4$ \\  $\mathrm{T_{d}}$ ($\mathrm{C_{2v}}$)}
    \end{subfigure}
    \hfill
    \begin{subfigure}{0.3\textwidth}
        \centering
        \includegraphics[width=0.7\textwidth]{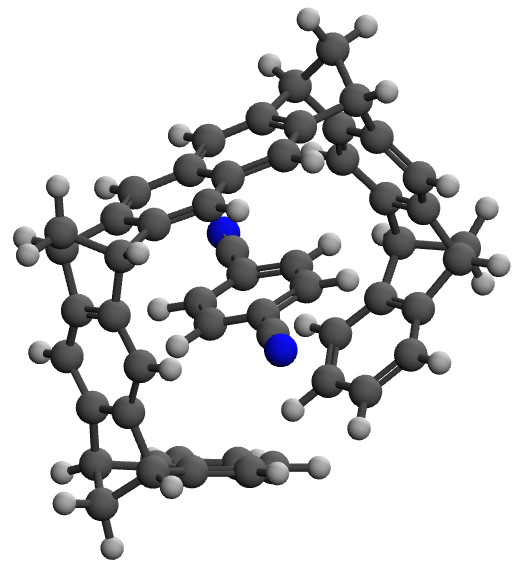}
        \caption{Tweezer \\  $\mathrm{C_{2}}$ ($\mathrm{C_{2}}$)}
    \end{subfigure}
    \vspace{0.5cm}
    \begin{subfigure}{0.3\textwidth}
        \centering
        \includegraphics[width=0.7\textwidth]{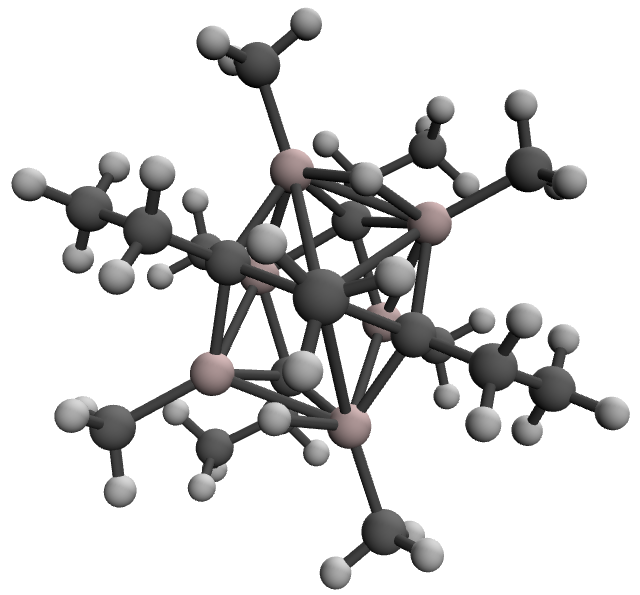}
        \caption{Carbaalane 1 \\ $\mathrm{C_{1}}$ ($\mathrm{C_{1}}$)}
    \end{subfigure}
    \hfill
    \begin{subfigure}{0.3\textwidth}
        \centering
        \includegraphics[width=0.7\textwidth]{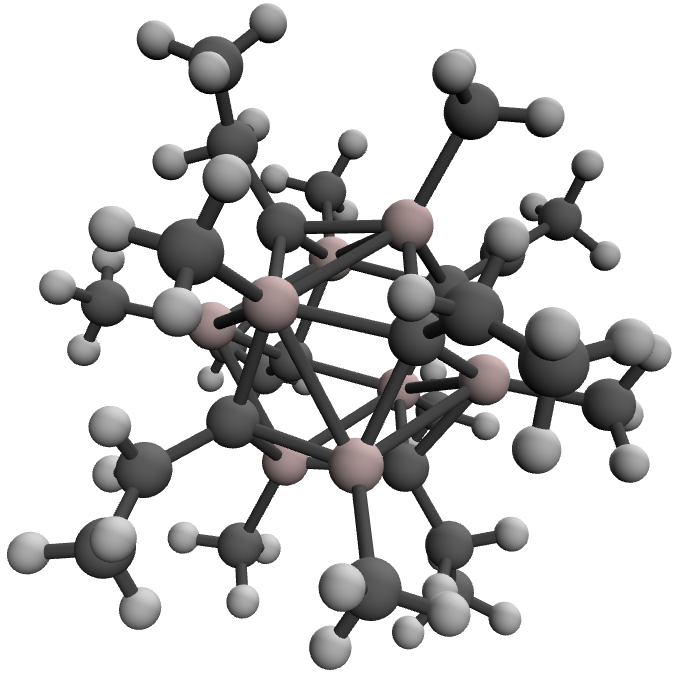}
        \caption{Carbaalane 2 \\  $\mathrm{C_{1}}$ ($\mathrm{C_{1}}$)}
    \end{subfigure}
    \hfill
    \begin{subfigure}{0.3\textwidth}
        \centering
        \raisebox{2cm}{%
        \includegraphics[width=0.9\textwidth]{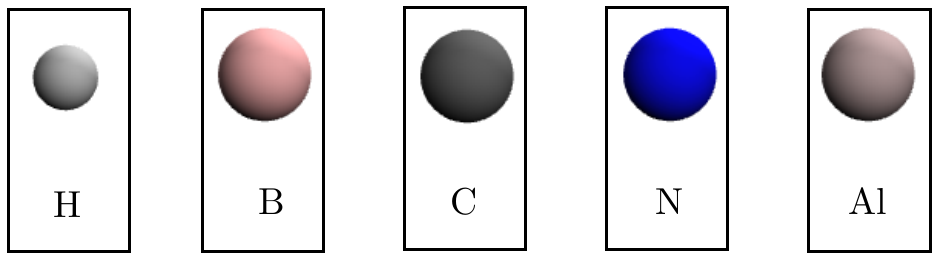}}
    \end{subfigure}
    \caption{Structures of buckminsterfullerene C$_{60}$, hexabenzocoronene (HBC), tetrameric cyclopentadienyl aluminum(I), coronene, tetrakis\-(t-butyl)tetraborane, a tweezer host-guest complex (tweezer), (AlMe)$_7$(CCH$_2$Me)$_4$H$_2$ (carbaalane 1) and (AlMe)$_8$(CCH$_2$Me)$_5$H (carbaalane 2). The full molecular point group is listed under the corresponding name, with the computational point group in parentheses.}
    \label{fig:NMR_systems}
\end{figure}

To demonstrate the applicability of the CD-GIAO-MP2 scheme for computing NMR shielding tensors and chemical shifts, calculations on various large molecules involving up to 100 atoms and more than 1000 basis functions are presented. The set of molecules includes buckminsterfullerene C$_{60}$, coronene C$_{24}$H$_{12}$, hexabenzocoronene (HBC) C$_{42}$H$_{18}$, tetrameric cyclopentadienyl aluminum(I) Al$_4$Cp$_{4}$, tetrakis(t-butyl)tetrabo\-rane {B$_4$tBu$_4$}, a tweezer host-guest complex C$_{54}$N$_{2}$H$_{36}$, and two carbaalanes (AlMe)$_7$(CCH$_2$Me)$_4$H$_2$ (\textit{closo}-structure) and (AlMe)$_8$(CCH$_2$Me)$_5$H (\textit{arachno}-structure). The structures and point-group symmetry of these molecules can be found in Fig.~\ref{fig:NMR_systems}.
In the case of B$_4$tBu$_4$ and Al$_4$Cp$_{4}$, the geometries were taken from Ref.~\onlinecite{Kollwitz98}, in the case of the carbaalanes from Ref.~\cite{Uhl01} while the geometries of HBC and C$_{60}$ were taken from Refs.~\onlinecite{Ochsenfeld01} and \onlinecite{Haeser89}, respectively.
Polarization functions from Ref.~\onlinecite{Gauss93} and versions of the Karlsruhe basis sets\cite{Schaefer92} dzp\footnote{4s1p/2s1p for H, 8s4p1d/4s2p1d for B, C, N, O, 11s7p1d/6s4p1d for Al} and tz2p\footnote{5s2p/3s2p for H, 9s5p2d/5s3p2d for B, C, N, O, 12s9p2d/7s5p2d for Al} have been used to calculate the shielding tensors of all molecules except the tweezer host-guest complex for which the smaller tzp\footnote{5s1p/3s1p for H, 9s5p1d/5s3p1d for C, N} basis was used instead of tz2p.
Both for the calculations exploiting point-group symmetry and the calculations with no point-group symmetry, Table~\ref{tab:TimingsMemory_NMR} reports the number of electrons and basis functions $N_{\mathrm{el}}$ and $N_{\mathrm{bf}}$, the timings for the CD $t_{\mathrm{CD}}$, the overall timings of the calculations $t_{\mathrm{total}}$ as well as the memory requirements. \\
The results clearly show a significant speed-up in the total timings of the NMR shielding calculation when exploiting molecular point-group symmetry. For HBC, the calculation is about 4.9 times faster when symmetry is used. As a general rule, calculations become more efficient with increasing order $h$ of the computational point group. This is illustrated by comparing coronene ($h$=8) and Al$_4$Cp$_{4}$ ($h$=4): a speed-up of about 4.2 is observed for coronene, compared to only 2.9 for Al$_4$Cp$_{4}$. 
Interestingly, while enforcing point-group symmetry provides remarkable savings in the overall calculations, the determination of the CVs is itself more expensive when using symmetry-adapted linear combinations (SALCs) of atomic orbitals. This is due to the fact that SALCs, contrary to AOs, are more delocalized, which not only makes Cauchy-Schwarz screening less effective, increasing thus the cost of the integrals evaluation, but also create a much larger number of non-negligible diagonal elements, making also the first step of the CD procedure more expensive, though the resulting number of CVs is not so much different. 
Nevertheless, exploiting symmetry in the subsequent steps of the calculation is so valuable that even for C$_{60}$ with a double-zeta basis set, where the CD is the most expensive step of the overall calculation, enforcing symmetry remains worth the effort, as can be seen by looking at the total elapsed times reported in Table~\ref{tab:TimingsMemory_NMR}. 
A reduction in the memory requirements by about the order of the molecular point group is expected and confirmed in the results. In case of Al$_4$Cp$_{4}$, for example, the computational point group is $\mathrm{D_{2}}$ with $h=4$, and both for dzp and tz2p, the memory requirements are reduced by a factor of 3.8. This trend also holds for higher-order point groups: for coronene and HBC in computational point group $\mathrm{D_{2h}}$, memory requirements are about 7 times lower. \\

\begingroup
\begin{table} [H]
\scriptsize
    \centering 
    \caption{Computational cost for the CD-GIAO-MP2 NMR shielding calculations of the molecules depicted in Fig.~\ref{fig:NMR_systems}. The table contains the number of basis functions, electrons, and Cholesky vectors ($N_{\mathrm{el}}$, $N_{\mathrm{bf}}$, and $M$), timings for the CD $t_{\mathrm{CD}}$ (in s), total timings $t_{\mathrm{total}}$ (h:min) and virtual memory VM (in GB). Superscript {\it sym} indicates calculations exploiting molecular point group symmetry, while superscript {\it nosym} indicates calculations without exploiting point group symmetry. 
    All calculations but the ones marked with an asterisk have been performed on a dual Intel Xeon Gold 6438Y+ CPU node with a total of 64 cores $@$ 2 GHz (4 GHz turbo boost) equipped with 512 GB of memory using 32 Cores and a Cholesky threshold $\tau$ of $10^{-5}$. The calculations marked with an asterisk have been performed on a quad Xeon Gold 6418H CPU node with a total of 192 $@$ 2.1 GHz (4 GHz turbo boost) equipped with 4 TB of memory, always using 32 cores.}
    \begin{tabularx}{\textwidth}{>{\centering\arraybackslash}p{0.150\textwidth} cccccccccccc}
        \toprule
        Molecule & $N_{\mathrm{el}}$ & Basis & $N_{\mathrm{bf}}$ & $t_{\mathrm{CD}}^{\mathrm{sym}}$ & $t_{\mathrm{CD}}^{\mathrm{nosym}}$ & $t_{\mathrm{total}}^{\mathrm{sym}}$ & $t_{\mathrm{total}}^{\mathrm{nosym}}$ & VM$^{\mathrm{sym}}$ & VM$^{\mathrm{nosym}}$ & $M^{\mathrm{sym}}$ & $M^{\mathrm{nosym}}$ \\
        \midrule
        \multirow{2}{*}{C$_{60}$} & \multirow{2}{*}{360} & dzp & 900 & 4028 & 400 & 01:31 & 03:29 & 42.78 & 325.26 & 7552 & 6058 \\
        &  & tz2p$^*$ & 1440 & 12792 & 1429 & 04:33 & 16:02 & 137.62 & 1037.59 & 10558 & 8442 \\ [5pt]
        \multirow{2}{*}{HBC} & \multirow{2}{*}{270} & dzp & 720 & 349 & 197 & 00:14 & 01:19 & 18.30 & 130.10 & 5538 & 4773 \\
        &  & tz2p & 1170 & 1140 & 634 & 00:47 & 05:07 & 60.80 & 435.80 & 8048 & 6950 \\ [5pt]
        \multirow{2}{*}{Al$_4$Cp$_4$} & \multirow{2}{*}{192} & dzp & 492 & 142 & 58 & 00:06 & 00:18 & 8.62 & 32.67 & 3644 & 3114 \\
        &  & tz2p & 788 & 498 & 191 & 00:19 & 01:06 & 28.49 & 107.84 & 5193 & 4509 \\ [5pt]
        \multirow{2}{*}{Coronene} & \multirow{2}{*}{156} & dzp & 420 & 88 & 44 & 00:03 & 00:09 & 2.55 & 17.79 & 3257 & 2778 \\
        &  & tz2p & 684 & 271 & 136 & 00:09 & 00:36 & 8.76 & 61.60 & 4720 & 4051 \\ [5pt]
        \multirow{2}{*}{B$_4$tBu$_4$} & \multirow{2}{*}{152} & dzp & 480 & 90 & 55 & 00:05 & 00:12 & 6.58 & 24.72 & 3626 & 3142 \\
        &  & tz2p & 804 & 292 & 210 & 00:14 & 00:55 & 24.47 & 91.23 & 5409 & 4709 \\ [5pt]
        \multirow{2}{*}{Tweezer} & \multirow{2}{*}{374} & dzp & 1020 & 544 & 551 & 02:08 & 05:48 & 238.47 & 469.24 & 7257 & 6733  \\
        &  & tzp$^*$ & 1280 & 763 & 1258 & 04:02 & 22:03 & 412.16 & 805.69 & 7965 & 7171 \\ [5pt]
        \multirow{2}{*}{Carbaalane 1} & \multirow{2}{*}{157} & dzp & 661 &  \multicolumn{2}{c}{123} &  \multicolumn{2}{c}{00:42}  &  \multicolumn{2}{c}{93.22} & \multicolumn{2}{c}{4144} \\
        &  & tz2p & 1067  & \multicolumn{2}{c}{448}  &  \multicolumn{2}{c}{03:01}  & \multicolumn{2}{c}{308.48} & \multicolumn{2}{c}{6052} \\ [5pt]
        \multirow{2}{*}{Carbaalane 2} & \multirow{2}{*}{208} & dzp & 834 & \multicolumn{2}{c}{252} & \multicolumn{2}{c}{01:53} &  \multicolumn{2}{c}{221.78} & \multicolumn{2}{c}{5202} \\
        &  & tz2p$^*$ & 1348 & \multicolumn{2}{c}{1652} &  \multicolumn{2}{c}{15:26}  &  \multicolumn{2}{c}{726.71} & \multicolumn{2}{c}{7561} \\
        \bottomrule 
    \end{tabularx}
    \label{tab:TimingsMemory_NMR}
\end{table}
\endgroup
\normalsize  

\begin{figure}[H]
    \centering
    \begin{subfigure}{0.49\textwidth}
        \centering
        \includegraphics[width=0.99\textwidth]{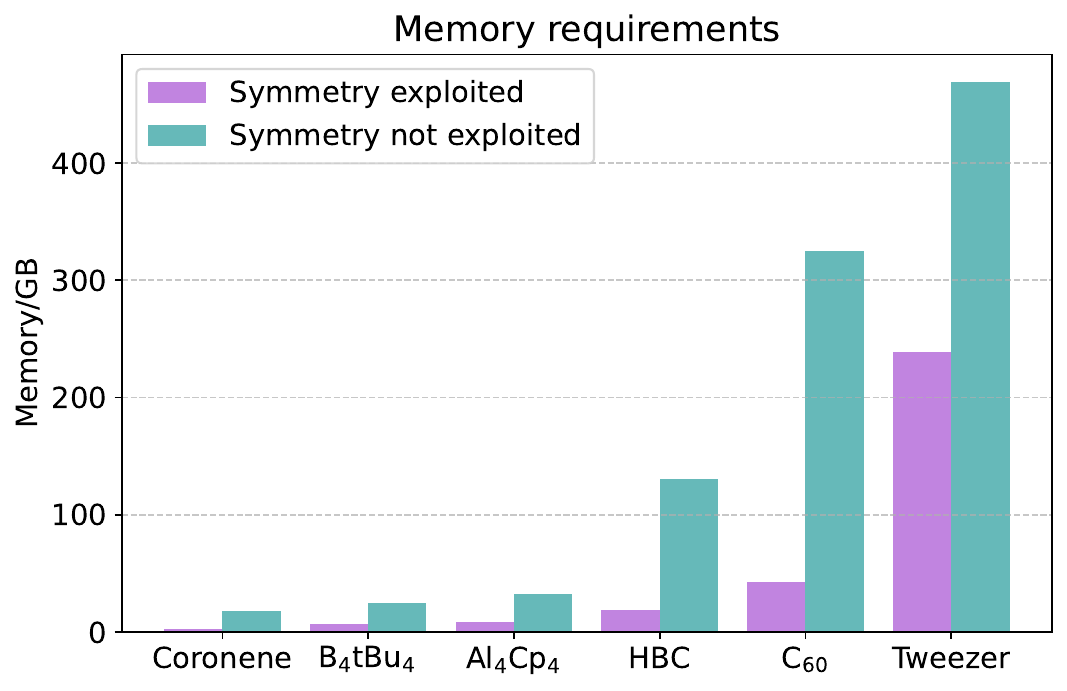}
    \end{subfigure}
    \hfill
    \begin{subfigure}{0.49\textwidth}
        \centering
        \includegraphics[width=0.99\textwidth]{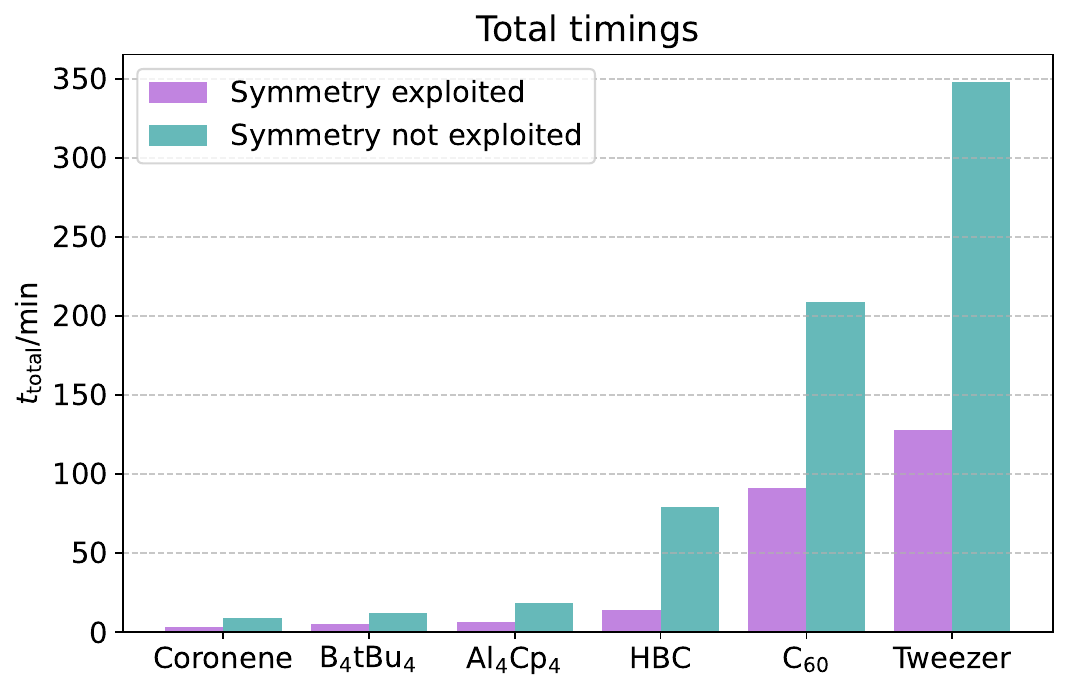}
    \end{subfigure}
    \caption{Comparison of the memory requirements and total timings for the NMR shielding calculations (dzp basis) performed with and without exploitation of molecular point-group symmetry.}
    \label{fig:Compare_sym_nosym}
\end{figure}

In Table~\ref{tab:MagnetizCoronene}, we report results for the magnetizability tensor of coronene as already published in Ref.~\onlinecite{Burger25}. The calculation was performed at the HF level using the tz2p basis set\cite{Schaefer92} and a Cholesky threshold $\tau$ of $10^{-6}$. It should be noted that these calculations were performed using the less efficient one-step scheme and that molecular point-group symmetry was exploited. 
\\
Although these are only HF results, the computation of the magnetizability tensor for systems with more than 1000 basis functions becomes quite demanding. Computing magnetizabilities is more elaborate than computing NMR shieldings since more terms need to be dealt with and concerning the CD of the integrals, three types of CVs (unperturbed, perturbed and doubly perturbed CVs) are necessary. 
\begin{table} [H]
    \centering
    \caption{Reported are the total timings $t_{\mathrm{total}}$ and timings for the CD $t_{\mathrm{CD}}$, as well as the values for the non-vanishing components of the magnetizability tensor $\xi_{ij}$ and isotropic magnetizability $\xi_{\mathrm{iso}}$ (in a.u.) for coronene computed using CD-GIAO-HF/tz2p and a Cholesky threshold of $10^{-6}$. The calculation has been performed with 8 cores on an Intel(R) Xeon(R) CPU E5-2643 v3 $@$ 3.40 GHz node.}
    \begin{tabular}{cccccc}
        \toprule
        $\xi_{xx}$ & $\xi_{yy}$ & $\xi_{zz}$ & $\xi_{\mathrm{iso}}$ & $t_{\mathrm{CD}}$/s & $t_{\mathrm{total}}$/h:min \\
        \midrule
        -25.083 & -25.088 & -130.301 & -60.157 & 14583 & 5:08 \\
        \bottomrule
    \end{tabular}
    \label{tab:MagnetizCoronene}
\end{table}
Ongoing work focuses on the use of the two-step algorithm for the CD of the magnetizability integrals and on the extension of the capability of \cfour\ to compute magnetizability tensors at MP2, CC, and CASSCF levels of theory.

We also mention that CD based NMR chemical shift computations at HF, MP2, and CASSCF level can be used to generate the required input files 
(containing unperturbed and perturbed density matrices) for the determination of magnetically induced current densities.\cite{Juselius04,Sundholm16}
Recent examples for applications can be found in Ref.~\onlinecite{Wang24} and \onlinecite{Wang26}. CD based NMR chemical shift computations have also recently been used within a QM/MM framework\cite{Senn09,Kirsch22} to study the gas to liquid shift of water.\cite{Burger25a}

\subsection{Linear response with CASSCF using Cholesky decomposition}
\label{section8}
CD has also been extensively exploited to accelerate the calculation of response properties at the CASSCF level of theory. Specifically, \cfour~ provides the capability to compute static properties, such as polarizabilities and NMR shielding tensors, as well as frequency-dependent (FD) properties, such as optical rotation.\cite{Nottoli22b,Nottoli25b} In addition, linear-response excitation energies are available.

The determination of all of these properties requires repeated operations involving the CASSCF Hessian matrix and, for FD properties and excitation energies, the metric matrix. In our implementation, the action of the Hessian (or metric) on a trial vector is evaluated directly, thereby avoiding the explicit construction and storage of these matrices. The orbital contribution to these matrix-vector products requires the evaluation of modified Fock matrices built either from one-index-transformed integrals or from transition density matrices. In both cases, the use of CD substantially accelerates the computation, reducing the leading computational cost to $\mathcal{O}(N^2AM)$ and $\mathcal{O}(N^2OM)$, respectively.

Besides their computational cost, the CASSCF response equations are notoriously challenging because of the ill-conditioning of the underlying matrices. To address this issue, we developed the swapped-metric orthogonal generalized Davidson (SMO-GD) algorithm.\cite{Alessandro23} 
The method employs metric-orthogonal expansion vectors, simplifying the projected problem within the reduced subspace and yielding a projected eigenvalue problem of half the dimension of the corresponding generalzied problem. Furthermore, the solver relies on numerically stable and robust orthogonalization procedures that generate vectors orthogonal to a prescribed subspace -- and mutually orthogonal -- to machine precision.

\begin{figure}[htb]
    \centering
    \includegraphics[width=0.9\linewidth]{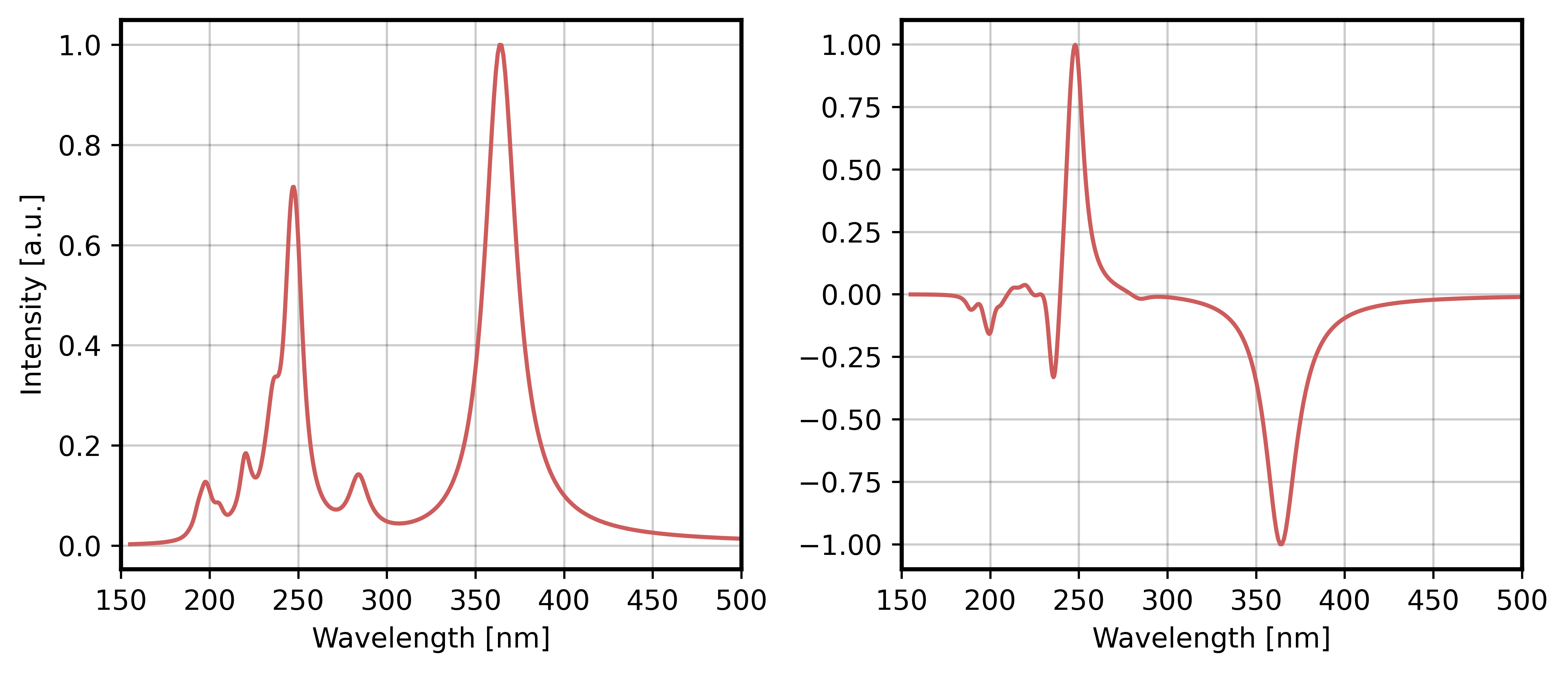}
    \caption{UV/VIS (on the left) and ECD (on the right) spectra of the anti-syn-anti conformation of phycocyanobilin computed at the LR-CASSCF(12,12)/def2-TZVP level of theory. The transitions were dressed with Lorentzian functions using a half-width at half-maximum of 0.1, while the intensities were normalized with respect to the maximum value.}
    \label{fig:bilin_asa_spectra}
\end{figure}

\begin{figure}[htb]
    \centering
    \includegraphics[width=0.45\linewidth]{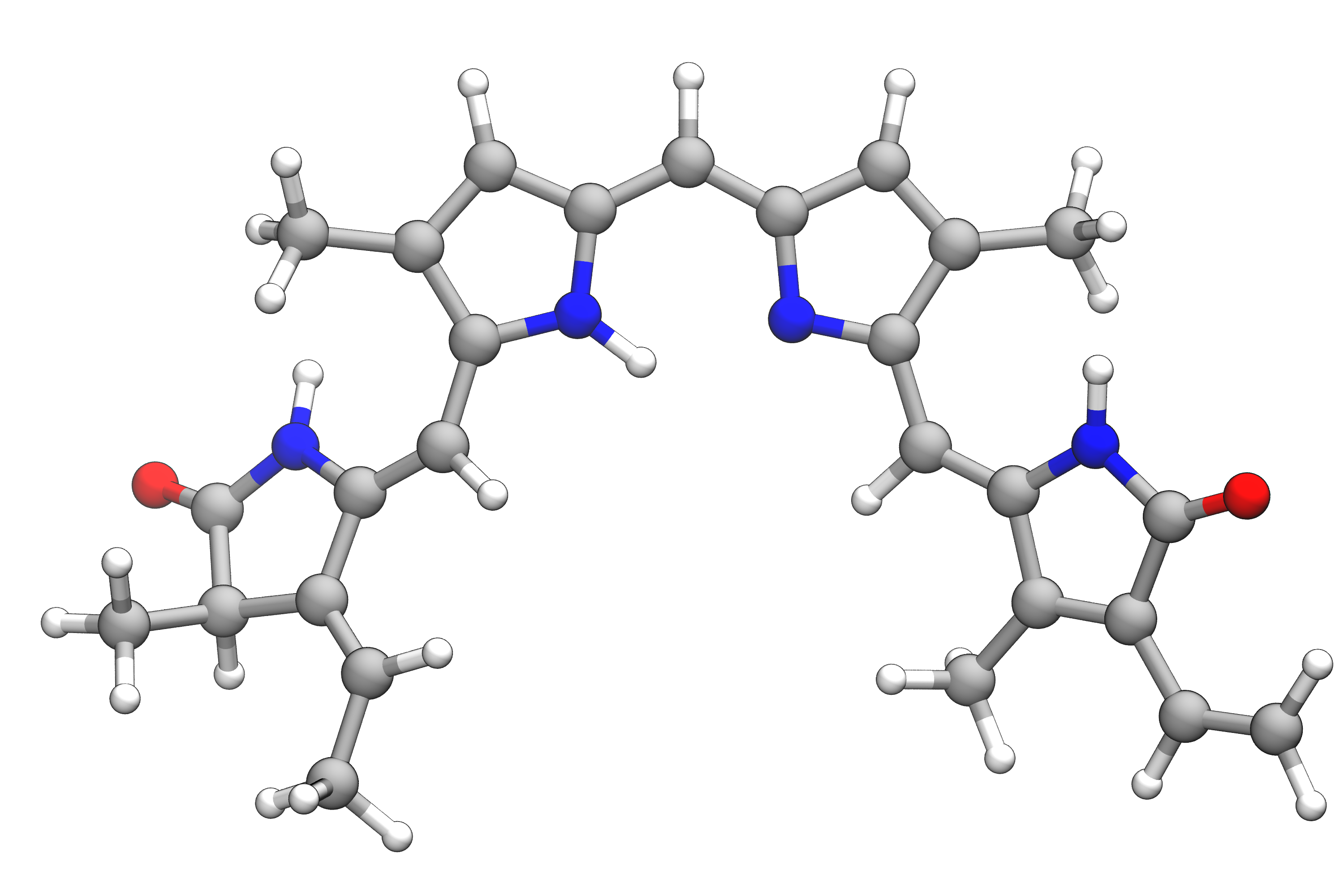}
    \caption{Molecular representation of phycocyanobilin in the anti-syn-anti conformation.}
    \label{fig:bilin_asa}
\end{figure}

The efficiency and robustness of the implementation were assessed by computing UV/VIS and electronic circular dichroism spectra, together with the optical rotation dispersion, of a large molecular system using extended basis sets and targeting up to 100 excited states. The calculations were performed on a model structure derived from phycocyanobilin, in which the two propionate side chains were replaced by hydrogen atoms. Fig.~\ref{fig:bilin_asa} shows the model structure of phycocyanobilin used in the calculations, in its anti-syn-anti conformation. The calculation targeted 20 excited states and employed the def2-TZVP basis set,\cite{Weigend05} corresponding to 1378 basis functions, a (12,12) active space, and was carried out on a node equipped with two AMD EPYC 7282 16-cores processors using 32 OpenMP threads. In Fig.~\ref{fig:bilin_asa_spectra}, we report representative UV/VIS and electronic circular dichroism (ECD) spectra.
The total wall-time required for the whole calculation was about 26.1~h. Of this, 4.5~min were required for the Cholesky decomposition, 1.4~h on the SCF step (using the unrestricted natural orbital (UNO) procedure, which requires three additional UHF calculations), 1.3~h on the CASSCF optimization, and 23.4~h on the LR calculation. We note that the iterative solution of the LR equations required 28 iterations to converge 20 excited states to a root-mean-square residual norm below 10$^{-5}E_h$, using a trial subspace of 30 vectors. This corresponds to an average wall time of approximately 54~s per matrix--vector product involving the full CASSCF Hessian including its orbital, configurational, and mixed blocks, demonstrating the feasibility of targeting several roots for systems of this size using extended basis sets. 

\section{Relativistic calculations and Cholesky decomposition}
\label{section9}
\cfour\ also offers a range of schemes for relativistic quantum-chemical treatments,\cite{Stopkowicz08,Stopkowicz11a,Cheng11a,Cheng11b,Cheng14,Lipparini16,Liu18,Kirsch19,Liu21,Uhlirova24} intended in particular for calculations on heavy-element compounds.
\cite{Cheng17,Zhang22} These calculations, when performed at the CC level, are more expensive than corresponding non-relativistic calculations and thus the use of techniques such as CD are here at least as important. However, the actual challenges differ for the various schemes. For the straightforward four-component approaches,\cite{Grant84,Visscher94,Saue97,Visscher95,Visscher1996,Jensen96,Fleig07,Thyssen08,Nataraj10} additional ERIs are needed and render the calculations expensive and quite demanding in terms of memory.
The increased computational cost in the two-component schemes with the inclusion of spin-orbit (SO) coupling, preferably those based on the exact two-component (X2C) decoupling scheme, \cite{Dyall1997,Kutzelnigg05,Ilias07,Liu2009}
can be attributed to the loss of spin symmetry in the ERIs in the MO representation. \cite{Liu21} 
We mention that spin-free computations at the X2C level\cite{Cheng11b} only face the same issues as non-relativistic computations and CD can be applied in the same way without further adaptations. In the following, we discuss how CD is used in \cfour\ to speed up relativistic quantum-chemical calculations. We start with a discussion concerning the two-component methods. This subsection is then followed by a discussion concerning the use of CD in four-component methods.

\subsection{Cholesky-decomposition based relativistic two-component coupled-cluster methods} 

We have recently implemented the CD-based exact-two-component (X2C)-CCSD 
and EOM-CCSD methods\cite{Zhang24} together with the corresponding analytic schemes for the evaluation of first-order properties.\cite{Zhang25}
Relativistic X2C-CC and EOM-CC methods \cite{Liu21}
enable rigorous treatments of both relativistic and electron-correlation effects.
The X2C schemes\cite{Dyall1997,Kutzelnigg05,Ilias07,Liu2009} in the spinor representation, i.e., with SO coupling included in the MOs,
offer non-perturbative treatments of SO coupling that are more accurate than the perturbative treatments.
Furthermore, the spinor representation allows
compact representations of wave functions for
many low-lying electronic states of open-shell species containing heavy elements.\cite{Zhang22,Tufekci24,Bonar26}
These states are dominated by a single determinant in the spinor representation, while
they are multideterminantal in the representation of real-valued functions.
Therefore, spinor-based relativistic CC methods have a unique applicability for these challenging systems. 


The recent development of X2C theory to include relativistic two-electron contributions\cite{Wang25}
provides ``electrons-only'' Hamiltonians that can closely reproduce the accuracy of the parent four-component Hamiltonian. 
These practical X2C schemes 
efficiently include the relativistic two-electron contributions by exploiting the local nature 
of the relativistic two-electron interactions.\cite{Liu18,Knecht22,Zhang22a} Of practical importance is that these X2C schemes work with the untransformed two-electron Coulomb interaction; i.e.,
they share the same two-electron AO integrals as the corresponding non-relativistic calculations.
Therefore, the implementation of the X2C-CC and EOM-CC methods uses the same AO Cholesky vectors as the non-relativistic treatment. 

The computational overhead of the X2C-CC calculations comes from the spin-symmetry breaking;\cite{Visscher1996,Liu18b} the molecular spinors cannot be separated into the ones with $\alpha$ or $\beta$ spin. Consequently, the number of the molecular spinors is twice the number of the $\alpha$ or $\beta$ MOs. The size of the Cholesky vectors in the molecular spinor representation $L_{pq}^P$ thus is much larger than that of the Cholesky vectors in the AO representation $L_{\mu\nu}^P$. 
The molecular spinor two-electron integral matrices $\langle ab||ci\rangle$  and $\langle ab||cd\rangle$ with three or four virtual spinor labels
are also much larger in size than the AO two-electron integral matrix $\langle \mu\nu||\sigma \rho \rangle$. 
The implementations of CD-based non-relativistic CC methods use the Cholesky vectors in the MO representation (see section \ref{section5}). They explicitly construct the MO ERIs on the fly for the iterative solution of the CC equations. 
In contrast, it is essential to exploit AO-based algorithms for an efficient implementation of the X2C-CCSD 
and EOM-CCSD methods,\cite{Zhang24}
because AO-based algorithms avoid explicit calculations of the large $\langle ab||cd\rangle$ and $\langle ab||ci\rangle$ integral matrices. 

Let us take as an example the particle-particle ladder term. This term can be rewritten in the AO-based algorithm as
\begin{eqnarray}
\frac{1}{2}\sum_{e,f}\langle ab||ef\rangle t_{ij}^{ef} =
\frac{1}{2} \sum_{\mu, \nu, \sigma, \rho} C^\ast_{\mu a}C^\ast_{\nu b}  \{\langle \mu\nu||\sigma \rho \rangle [ \sum_{e,f} t_{ij}^{ef} C_{\sigma e}C_{\rho f}] \},
\end{eqnarray}
in which $C$ represents the molecular spinor coefficients. Here the molecular spinor ERIs $\langle ab||ef\rangle$ and the cluster amplitudes $t_{ij}^{ef} $ are spin-dependent. On the other hand, the AO integrals $\langle \mu\nu||\sigma \rho \rangle$ remain spin-free. The spin factorization of the AO ERIs in the AO-based algorithm reduces the floating-point operation count by around a factor of four\cite{Liu18b} compared to the MO-based algorithm with explicit assembly of the $\langle ab||ef\rangle$ integrals. 

Relativistic two-component CC calculations often employ uncontracted basis sets, especially when targeting core properties that sample the electron density in the inner-shell region.
The use of uncontracted basis sets gives rise to a large number of high-lying virtual spinors.
The correlation of high-lying virtual spinors and inner-shell core spinors only makes small contributions to molecular properties.
Therefore, these spinors can be frozen safely in  relativistic CC and EOM-CC calculations.
On the other hand, it is necessary to treat the orbital-relaxation effects due to both frozen-core and frozen-virtual spinors for rigorous calculations
of analytic relativistic CC energy derivatives. 
This necessitates calculations of ERIs involving these frozen spinors
and accumulation of the corresponding contributions. An important computational benefit for the CD-based implementation of analytic relativistic CC first-order properties is the dramatic reduction of the computational cost associated with the calculation and processing of the additional integrals involving these frozen spinors.\cite{Zhang25} 

\begin{figure} [H]
    \centering
    \includegraphics[width=12cm]{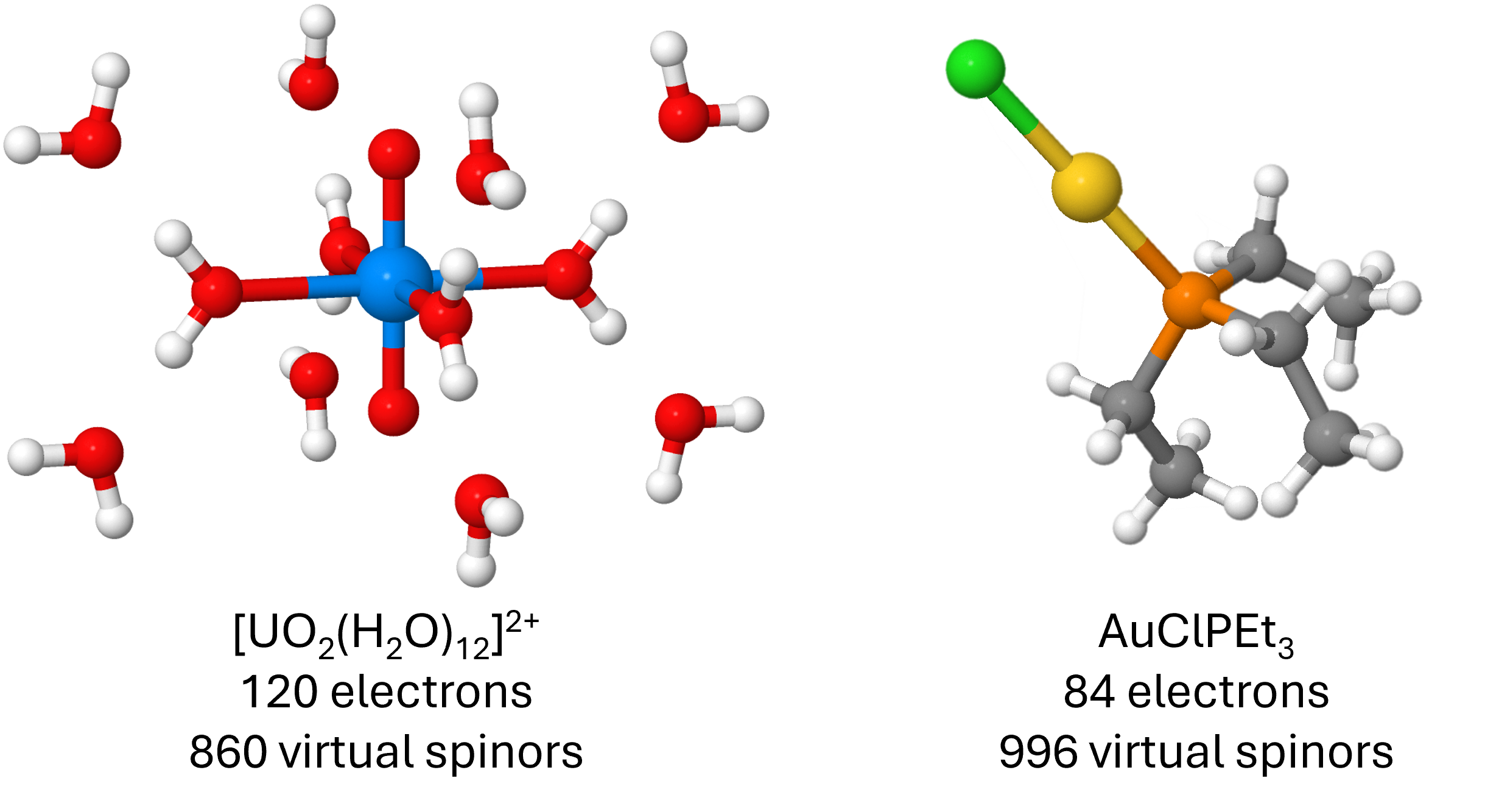}
    \caption{[UO$_2$(H$_2$O)$_{12}$]$^{2+}$ as an example of relativistic EOM-CCSD calculations for excited states and AuClPEt$_3$ as an example of relativistic CCSD first-order properties.} 
    \label{cd-2ccc}
\end{figure}

We have integrated the implementation of analytic CD-X2C-CCSD 
and EOM-CCSD first-order properties with the X2C property-integral module in \cfour\ to enable the calculations of a variety of first-order properties. 
Apart from electric and magnetic parameters, the interface with the X2C property-integral module allows efficient calculations of the symmetry-violating sensitivity parameters\cite{Zhang21a,Chen24} pertinent to the searches of new physics beyond the Standard Model by means of precision measurement of polar molecules containing heavy atoms.\cite{Safronova18,Cairncross19,DeMille24}

The CD-based implementation significantly improves the efficiency of relativistic CC and EOM-CC calculations.
X2C-CCSD and EOM-CCSD calculations that correlate around 100 electrons and 1000 virtual spinors can now routinely be carried out.
For example, X2C-EOM-CCSD calculations for the low-lying excited states in the uranyl ion with two shells of solvent molecules, [UO$_2$(H$_2$O)$_{12}$]$^{2+}$, 
correlated 120 electrons and 860 virtual spinors (Fig.~\ref{cd-2ccc}). SO coupling makes significant contributions to the excitation energy of the lowest excited state, reducing the excitation energy by around 0.4 eV.\cite{Zhang24}
The usefulness of the CD-based analytic X2C-CCSD first-order properties has been demonstrated in calculations of the M{\"o}ssbauer spectra for gold complexes,\cite{Zhang25} in which the obtained X2C-CCSD results for contact densities and electric-field gradients correlate well with the measured isomer shifts and quadrupole splittings, respectively.  
The calculations for AuClPEt$_3$ 
correlated 84 electrons and 996 virtual spinors (Fig.~\ref{cd-2ccc}). The efficient CD-based implementation has thus extended the applicability of relativistic two-component CCSD and EOM-CCSD methods to medium-sized molecules containing heavy atoms. 

\subsection{Cholesky-decomposition based relativistic four-component methods}

In four-component relativistic calculations,\cite{Dyall07}
the large and small component of the molecular spinors 
\begin{eqnarray}
\psi_p =  \left( \begin{array}{c} \psi_p^L \\ \psi_p^S \end{array}\right)
\end{eqnarray}
are explicitly determined (for example within four-component self-consistent-field
computations) and then, usually with the no-pair approximation,\cite{Sucher80} used within electron-correlation treatments. A convenient way to deal
with four-component relativistic treatments is the introduction of the so-called
pseudo-large component $\phi_p^L$ to represent the small component $\psi^S_p$:\cite{Kutzelnigg84a,Dyall1997}
\begin{eqnarray}
\psi_p^S = \frac{1}{2c} {\bf \sigma} \cdot  {\bf p}  \phi_p^L.
\end{eqnarray}
This is in particular advantageous, as the large and pseudo-large components
can be expanded in the same AO basis (unlike the large and small component, where different basis
sets are required\cite{Dyall07} and attention has to be given to the issue of kinetic balance\cite{Stanton84}).
A detailed discussion of the formulation of four-component methods
in this framework can be, for example, found in Ref.~\onlinecite{Uhlirova24}.

\begin{figure}[H]
    \centering
        \includegraphics[width=0.6\textwidth]{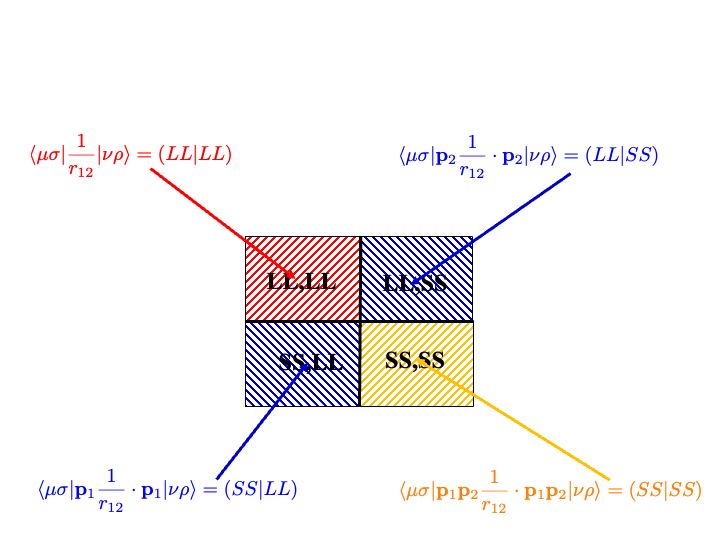}
    \caption{Arrangement of the spin-free relativistic ERIs in a positive semidefinite ERI matrix}
    \label{rel4c1}
\end{figure}

The following discussion 
focuses first on spin-free Dirac-Coulomb treatments, in which the SO part of the Hamiltonian is skipped.\cite{Dyall94} and only the instantaneous two-electron interactions are considered
For such treatments, the following ERIs need to be computed and processed besides the usual ERIs
\begin{eqnarray}
  (\psi^L_p \Psi^L_q | \frac{1}{r_{12}}| \psi^L_r \psi_s^L);   \nonumber \\
  (\psi^L_p \Phi^L_q | {\bf p}_2 \frac{1}{r_{12}} {\bf p}_2 |\psi^L_r \phi_s^L); \nonumber \\
  (\phi^L_p \Psi^L_q | {\bf p}_1 \frac{1}{r_{12}} {\bf p}_1 |\phi^L_r \psi_s^L);\nonumber \\
(\phi^L_p \Phi^L_q | {\bf p}_1{\bf p}_2 \frac{1}{r_{12}} {\bf p}_1{\bf p}_2 |\phi^L_r \phi_s^L)
\end{eqnarray}
which means that one has to evaluate the following ERIs in the given AO representation
\begin{eqnarray}
   (LL| LL) = \langle \mu \sigma|  \frac{1}{r_{12}} | \nu \rho\rangle; \nonumber \\  
    (LL| SS) = \frac{1}{14 c^2} \langle \mu \sigma| {\bf p}_2 \frac{1}{r_{12}} \cdot {\bf p}_2  | \nu \rho\rangle; \nonumber \\  
    (SS| LL) = \frac{1}{4 c^2} \langle \mu \sigma| {\bf p}_1 \frac{1}{r_{12}} \cdot {\bf p}_1  | \nu \rho\rangle; \nonumber \\  
    (SS| SS) =  \frac{1}{16 c^4} \langle \mu \sigma| {\bf p}_1 {\bf p}_2 \frac{1}{r_{12}} \cdot {\bf p}_1  {\bf p}_2| \nu \rho\rangle.
\end{eqnarray}
In the framework of CD, the question is how these additional integrals are decomposed.
A solution is obtained by sorting all integrals in a large positive semidefinite matrix
as shown in Fig.~\ref{rel4c1} and to apply CD to this expanded matrix instead the usual ERI matrix.\cite{Uhlirova24} The diagonal blocks of this matrix
contain the $(LL|LL)$ and $(SS|SS)$ integrals, while the 
$(SS|LL)$ and $(LL|SS)$ integrals appear in the off-diagonal blocks.
When performing a CD on this matrix, it is realized that due to the prefactor of $1/16c^4$ of the $(SS|SS)$ integrals the Cholesky basis is almost exclusively formed from the diagonals of the $LL,LL$ block
and that the final Cholesky basis contains, if at all, only a few functions from the $SS,SS$ block.\cite{Uhlirova24} This suggests that the selection of the Cholesky basis (within the two-step scheme) can be significantly simplified by limiting the selection to the $LL,LL$ block.\cite{Banerjee23,Uhlirova24} In particular, no $(SS|SS)$ integrals are then required for this step (which however would be the case if all diagonals are considered for the selection of the Cholesky basis). 

\begin{figure}[H]
    \centering
        \includegraphics[width=0.6\textwidth]{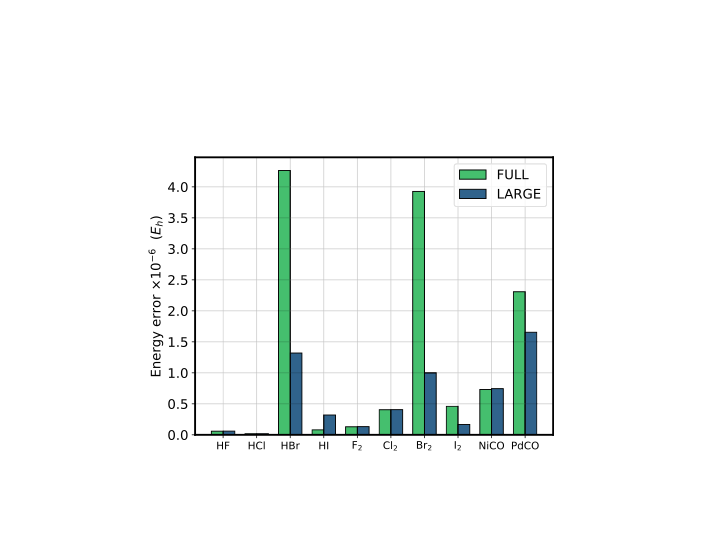}
    \caption{Error in the Cholesky based -SF Dirac-Coulomb CCSD energies (compared to energies obtained without CD) performed using the uncontracted ANO-RCC
    \cite{Roos04} 
    basis set using both the full Cholesky basis and those from just the $LL,LL$ block. The Cholesky tolerance has been set to 10$^{-5}$. Used with permission of ACS, from "Cholesky Decomposition in Spin-Free Dirac-Coulomb Coupled-Cluster Calculations", T. Uhl{\' i}{\v r}ov{\' a}, D. Cianchino, T. Nottoli, F. Lipparini, J. Gauss, {\sl J. Phys. Chem.} A, {\bf 128}, 8292–8303, 2024; permission conveyed through Copyright Clearance Center, Inc.
    }
    \label{rel4c2}
\end{figure}

Fig.~\ref{rel4c2} shows the differences in computed energies when using the full space or just the $LL,LL$
block for the selection of the Cholesky basis.
The observed differences are negligible in particular considering the error that already results from the CD
itself.

\begin{figure}[H]
    \centering
    \includegraphics[width=0.6\textwidth]{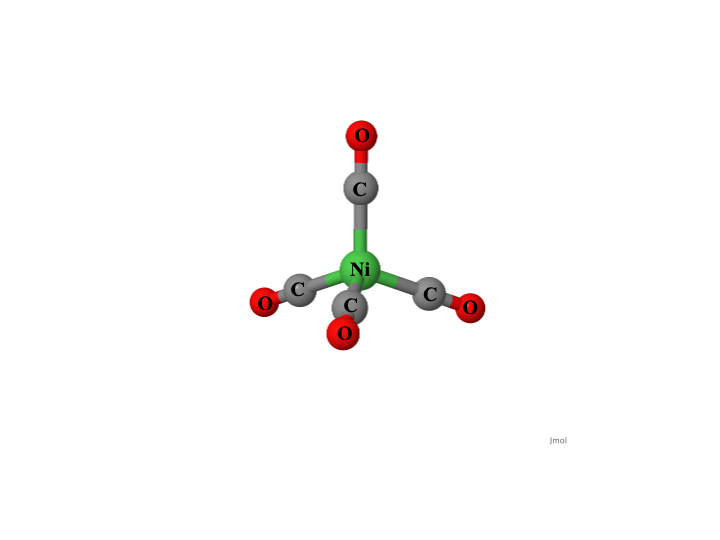}
    \caption{The Ni(CO)$_4$ molecule for which CD SFDC-CCSD computations using an uncontracted ANO-RCC basis (1016 basis functions)\cite{Roos04} have been performed.}
    \label{rel4c3}
\end{figure}

The CD leads then to Cholesky vectors (CVs) defined as
\begin{eqnarray}
{\bf L}^P = \left(  \begin{array}{c} {\bf L}^{L,P} \\ {\bf L}^{S,P} \end{array}\right)\nonumber \\
L^{L,P}_{\sigma \rho}= (\mu \nu | \nu \mu)^{-1/2} \left( (\sigma\rho| \nu \mu) - \sum_{R=1}^{P-1} L_{\sigma \rho}^{L,R} L_{\nu \mu}^{L,R} \right) \nonumber \\
L^{S,P}_{\sigma \rho}= (\mu \nu | \nu \mu)^{-1/2} \left( (\sigma\rho| {\bf p}_1 \frac{1}{r_{12}} {\bf p}_1 | \nu \mu) - \sum_{R=1}^{P-1} L_{\sigma \rho}^{S,R} L_{\nu \mu}^{L,R} \right).
\end{eqnarray}
The construction of the CVs thus does also not require the $(SS|SS)$ integrals; in the
calculation afterwards, these integrals are reconstructed from the $L^{S,P}$ CVs 
\begin{eqnarray}
(\sigma\rho| {\bf p}_1 {\bf p}_2 \frac{1}{r_{12}} |{\bf p}_1 {\bf p}_2 \nu \mu)\approx \sum_{P} L_{\sigma \rho}^{S,P} L_{\nu \mu}^{S,P}. 
\end{eqnarray}
The described CD procedure has been implemented in \cfour\ with the required ERIs computed using the \mint\ package. SF Dirac-Coulomb-CCSD (SFDC-CCSD) calculations are then easily possible with the CCSD code described in Section~\ref{section5}, as changes (apart from the use of a different, i.e., four-component, SCF code) are only required in the transformation of the Cholesky vectors from the AO to the MO basis. The transformation has now to be carried out according to 
\begin{eqnarray}
L_{pq}^P = \sum_{\mu, \nu} L_{\mu \nu}^{L,P} c_{\mu p}^L c_{\nu_q}^L + \sum_{\mu, \nu} L_{\mu \nu}^{S,P} c_{\mu p}^S c_{\nu_q}^S
\end{eqnarray}
with $c_{\mu p}^{L}$ and $c_{\mu p}^S$ as the MO coefficients of the large and small component, respectively. Large-scale SFDC-CCSD calculations on, for example, molecules such as Ni(CO)$_4$ (Fig.~\ref{rel4c3} and its heavier analogues are thus possible. 
For Ni(CO)$_4$, the SFDC-CCSD calculations using the  
uncontracted ANO-RCC basis set\cite{Roos04}
involve 1016 basis functions (13 frozen core orbitals), the error in the SFDC-HF and SFDC-CCSD energies compared to conventional calculations amounts to less than 10$^{-4}$ Hartree for a Cholesky threshold of $10^{-4}$ and 10$^{-6}$ Hartree for a threshold of $10^{-6}$. At the MP2 level, SFDC calculation on molecules such as auranofin (C$_{20}$H$_{34}$AuO$_9$PS, a molecule of interest because of its anticancer activity\cite{mirabelli85,Marzano07}) with
more than 1544 basis functions (uncontracted tzp basis\cite{Barbieri06} also are possible. For more details, see Ref.~\onlinecite{Uhlirova24} 

\begin{figure}[H]
    \centering
        \includegraphics[width=0.6\textwidth]{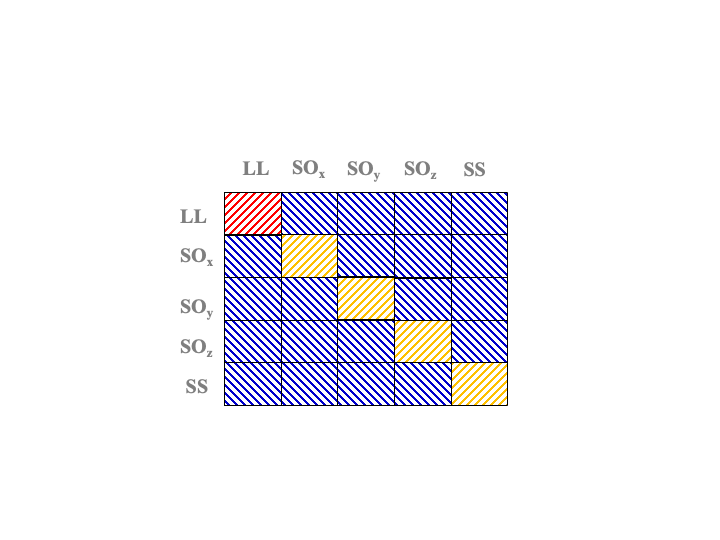}
    \caption{Arrangement of the relativistic ERIs including the SO integrals in a positive semidefinite ERI matrix.}
    \label{rel4c4}
\end{figure}

For full Dirac-Coulomb computations including SO interactions, the ERI matrix in four-component calculations look like in Fig.~\ref{rel4c4}
as we have no to consider in addition the SO integrals
$(LL|SO_i)$, $(SS|SO_i)$ and $(SO_i|SO_j)$ with $i= x,y,$ and $z$
\begin{eqnarray}
( LL| SO_i) = \frac{1}{4 c^2}  \langle \mu \sigma| ({\bf p_2}\frac{1}{r_{12}}\times {\bf p_2})_i| \nu \rho\rangle \nonumber \\ 
 ( SS| SO_i) = \frac{1}{16 c^4}  \langle \mu \sigma| ({\bf p}_1 ({\bf p}_2\frac{1}{r_{12}} \times{\bf p}_2)  {\bf p}_1)_i | \nu \rho\rangle =  ( SS| SO_i)
 \nonumber \\
 ( SO_i| SO_j) =  \frac{1}{16 c^4} \langle \mu \sigma| ({\bf p_1} \times( {\bf p_2} \times ( \frac{1}{r_{12}})) {\bf p_1} {\bf p_2} )_{i,j} | \nu \rho\rangle 
\end{eqnarray}
Again, one can limit the selection of the Cholesky basis to the $LL,LL$ block and then compute the Cholesky vectors as
\begin{eqnarray}
L^{X,P}_{\sigma \rho}= (\mu \nu | \nu \mu)^{-1/2} \left( (\sigma\rho| \nu \mu)^{\rm X,LL} - \sum_{R=1}^{P-1} L_{\sigma \rho}^{X,R} L_{\nu \mu}^{L,R} \right)\quad X = S, L, SO_i
\end{eqnarray}
$(SS|SS)$ integrals (now also the $(SS|SO_i)$ and $(SO_i| SO_j)$ blocks with $i=x, y,$ and $z$) are not needed and in the computation afterwards reconstructed from the appropriate Cholesky vectors. 
The current status in \cfour\ concerning CD based relativistic methods is that the CD and a CD based four-component SCF code is working. First results that demonstrate the applicability of CD in DC-HF computations are given in Table~\ref{tab1}.
\begin{table}
\begin{tabular}{|c|c|c|}
\hline
comp. scheme& E & \# CVs \\
\hline
CD, $\tau =10^{-4}$ &-2605.522558 &399 \\
CD,  $\tau =10^{-5}$ &-2605.522861 &461 \\
CD,  $\tau =10^{-6}$ & -2605.522851 & 520\\
no CD & -2605.522853 & --- \\
SFDC &-2605.442054 & --- \\
\hline
\end{tabular}  
\caption
{Computed energies (E in $E_h$) for HBr ($r$(HBr) = 2.672946 bohr) using an uncontracted cc-pVDZ basis\cite{Dunning89} at DC-HF level with and without CD. CD results are reported for different Cholesky thresholds $\tau$. For each calculation, the number of Cholesky vectors is also reported.}
\label{tab1}
\end{table}
Current work concerning the use of CD in four-component relativistic calculations focuses on the extension of the full Dirac-Coulomb treatments to MP2 and CC. However, unlike for the SF case, complex codes are here needed. A promising possibility is here to use for this purpose the available CC code in \qcumbre\cite{qcumbre} which is able to perform complex CC computations. Another topic of interest is the calculation of molecular properties using CD at the four-component level.

\section{Calculations in a finite magnetic field using Cholesky decomposition}
\label{section10}

In the case of finite magnetic-field (ff) calculations using \cfour, the electronic Hamiltonian is
\begin{align}
	\hat{H}_B&=\hat{H}_0+\hat{P}_\mathrm{orb}^O+\hat{P}_\mathrm{spin}+\hat{D}^O.
	\label{eq:mag_ham}
\end{align}
The additional three terms on top of the field-free electronic Hamiltonian $\hat{H}_0$ describe the interaction with the magnetic field $\bf{B}$. They are the paramagnetic orbital-Zeeman 
\begin{eqnarray}
    \hat{P}_\mathrm{orb}^O=\frac{1}
    {2}\sum\limits_{\alpha}^{N_{el}}\bf{B}\cdot\hat{\bm{l}}_{\alpha}^{\bf{O}}
    \end{eqnarray}
and the spin-Zeeman 
\begin{eqnarray}
\hat{P}_\mathrm{spin}=\sum\limits_{\alpha}^{N_{el}}\bf{B}\cdot\hat{\bf{s}}_{\alpha}
 \end{eqnarray}
 terms,
as well as the diamagnetic term 
\begin{eqnarray}
\hat{D}^O=\frac{1}{8}\sum\limits_{\alpha}^{N_{el}}\left[B^2{r}_{\alpha}^{\bf{O}^2}-\left(\bf{B}\cdot{\bf{r}}_{\alpha}^{\bf{O}}\right)^2\right]
\end{eqnarray}
with $\alpha$ as the electron index and $N_{el}$ as the number of electrons of the system.
While the spin-operator $\hat{\bf{s}}_{\alpha}$ only depends on the spin coordinates, the angular-momentum $\hat{\bf{l}}_{\alpha}^{\bf{O}} = -i ({\bf r}_{\alpha} - {\bf R}_O) \cross {\bf \nabla}_{\alpha}$ with origin ${\bf R}_O$ and the position ${\bf{r}}_{\alpha}^{\bf{O}} = {\bf r}_{\alpha} - {\bf R}_O$ operator involve the gauge origin ${\bf{R}}_O$ which is the arbitrary position where the vector potential 
vanishes, i.e., ${\bf{A}}({\bf{R}}_O)=\bf{0}$.
Similar to the calculation of magnetic properties,\cite{Hameka58,Ditchfield72,Wolinski90,Helgaker91,Gauss92} gauge-origin invariance is built into the quantum-chemical approximation by employing magnetic field-dependent basis functions, the so called GIAOs or London orbitals\cite{London37} (see Eq.~(\ref{giao}) in Section \ref{section7})
 In the ff case however, implications exist for the standard one- and two-electron integrals and not only for integral derivatives as in Section~{\ref{section7}}. Integrals over GIAOs are in general complex valued and specialized algorithms have been developed for their calculation in the last decades.\cite{Tellgren08,Tellgren12,Irons17,Pausch20,Irons21,Blaschke24} These algorithms have enabled the development of ff approaches for the treatment of arbitrary field strengths and field orientations.\cite{Tellgren14,Furness15,Stopkowicz15,Hampe17,Bischoff20,Lehtola20,Wibowo2021,LONDON,WilliamsYoung20,qcumbre,bagel,quest} The evaluation of integrals over GIAOs has been also implemented in the \mint\ module\cite{mint} of \cfour\ using the McMurchie-Davidson scheme.\cite{McMurchie78}

An important issue is that for the ERIs over GIAOs in the presence of a magnetic field, the eightfold permutational symmetry is in general reduced to fourfold\
\begin{equation}
	(\mu \nu | \sigma \rho) = ( \sigma \rho | \mu \nu ) = ( \nu \mu | \rho \sigma )^* = (  \rho \sigma | \nu \mu )^*.
\end{equation}
The index pairs associated to one electron $\mu\nu$ and $\sigma\rho$ can no longer be exchanged independently. 
This symmetry reduction directly affects the CD procedure. 
First, the diagonal elements of the Cauchy-Schwarz inequality that constitute the positive semidefiniteness of the ERI matrix in the complex case are
\begin{equation}
	 D_{\mu \nu} = (\mu \nu | \nu \mu) 
\end{equation}
which need to be used as pivoting elements. It is highlighted that in the real case (unlike for the complex case\cite{Blaschke22}) the $(\mu \nu | \mu \nu)$ elements are equal to $(\mu \nu | \nu \mu)$ elements and the latter
are usually chosen in the formulation of the CD.
Second, the fourfold symmetry cannot be accounted for by simply restricting the basis-set indices to $\mu \geq \nu$ and $ \sigma \geq \rho$. This restriction in the field-free case implies symmetric Cholesky vectors
\begin{equation}
	L_{\mu \nu}^P = L_{\nu \mu}^P.
\end{equation}
Instead, the Hermiticity of the ERIs leads to the following relationship for the Cholesky vectors
\begin{equation}
	L_{\mu \nu}^P = (L_{\nu \mu}^{P'})^*, \label{eq:CVherm}
\end{equation}
where index $P$ is associated to $\sigma\rho$ and $P'$ is associated to $\rho\sigma$.\cite{Gauss23}

It should be noted that the intricacies of ff calculations more than double the memory requirements as compared to a field-free calculation due the decreased permutational symmetry of the ERIs and the need to store complex rather than real numbers. The employment of complex algebra increases the number of the required floating-point operations by a factor of 4. Further considerations within specialized BLAS routines can reduce this prefactor to 3 in the case of matrix-matrix multiplications that are employed for tensor contractions.\cite{Dongarra90} Moreover, to account for the distortion of the orbitals due to the magnetic field, large uncontracted (unc) basis sets 
need to be employed.\cite{Farhaz25} Such large basis sets, however, often introduce linear dependencies that hinder the convergence of iterative procedures.   
The CD can potentially be very effective in tempering these implications as it reduces the memory requirements and the number of required floating-point-operations due to a rank reduction of the tensors involved.
In addition, it eliminates linear dependencies up to the Cholesky threshold improving convergence when large uncontracted basis sets are employed as is standard in ff calculations.\cite{Blaschke22,Gauss23,Blaschke24b}

The original implementation of the CD over GIAOs in the \mint\ package of \cfour\ has been reported in Ref.~\onlinecite{Blaschke22}. It was based on the partial pivoting algorithm and did not take into account the Hermiticity of the Cholesky vectors. Practically, this means that the relationship of Eq.~\eqref{eq:CVherm} is not ensured and that these elements have to be explicitly handled and stored. Later developments\cite{Gauss23} account for the Hermiticity relation by symmetrically constructing the elements $L_{\mu \nu}^{P\rightarrow (\sigma \rho)}$ and $L_{\mu \nu}^{P'\rightarrow (\rho \sigma)}$ during the iterations of the CD procedure. 
The exact steps that account for this property within the CD algorithm are presented in detail in Refs.~\onlinecite{Gauss23} and \onlinecite{Blaschke24c}. Memory savings can be achieved when the Hermiticity is not violated by only explicitly computing and storing the linearly independent elements of the Cholesky vector. Both options are available in the \cfour\ program, as well as a CD for the ff case using the two-step procedure.\cite{Blaschke24c}

The implemented CD in the case of ff calculations further benefits from the exploitation of molecular symmetry through point-group theory for both real and complex Abelian point groups.\cite{Kitsaras26}

A ff HF procedure based on Cholesky vectors has been implemented in \cfour, reducing the scaling of the calculation to $\mathcal{O} (MN^2O)$.
The ff SCF implementation follows the standard field-free implementation as described in previous sections. The MP2 approach based on Cholesky vectors for the ff case has been implemented as well, where the bottleneck of the transformation of the ERIs from the atomic- to molecular-orbital basis is reduced to $\mathcal{O}(N^3 M)$.\cite{Blaschke22} 
In addition, a ff first-order CASSCF implementation that exploits CD as well as point-group symmetry has been recently developed.\cite{Guidone25}
Calculations at the ff CC and ff EOM-CC levels of theory are possible via an interface to the \qcumbre\ program package.\cite{qcumbre,Hampe17,Hampe19,Hampe20,Kitsaras24,Blaschke24b,Blaschke24c,Kitsaras25,Grazioli25} As mentioned in previous sections, the formal scaling of a CC calclation is not reduced by the use of Cholesky decomposed integrals, but the scaling prefactor becomes smaller and the integral storage and memory-bandwidth requirements decrease. Details on the implementation for the ff-CD-(EOM)-CC2 and ff-CD-(EOM-)CCSD approximations in \qcumbre\ are found in Refs.~\onlinecite{Blaschke24b} and \onlinecite{Blaschke24c} accompanied by thorough discussions on the scaling of individual steps and bottlenecks. 

To demonstrate the utility of using CD in the case of ff calculations, two examples from previous studies are presented in the following. First, the methylidyne radical (CH) is examined at the ff-CD-MP2 level of theory for different orientations and strengths of the field in order to assess the effectiveness of the CD for ff calculations.\cite{Blaschke22} Second, the effect of a He atmosphere to a Mg electronic transition is studied at the ff-CD-EOM-CCSD and ff-CD-EOM-CC2 levels of theory.\cite{Blaschke24b}

\begin{figure}[H]
    \centering
\begin{subfigure}{0.45\textwidth}
    \includegraphics[width=1\textwidth]{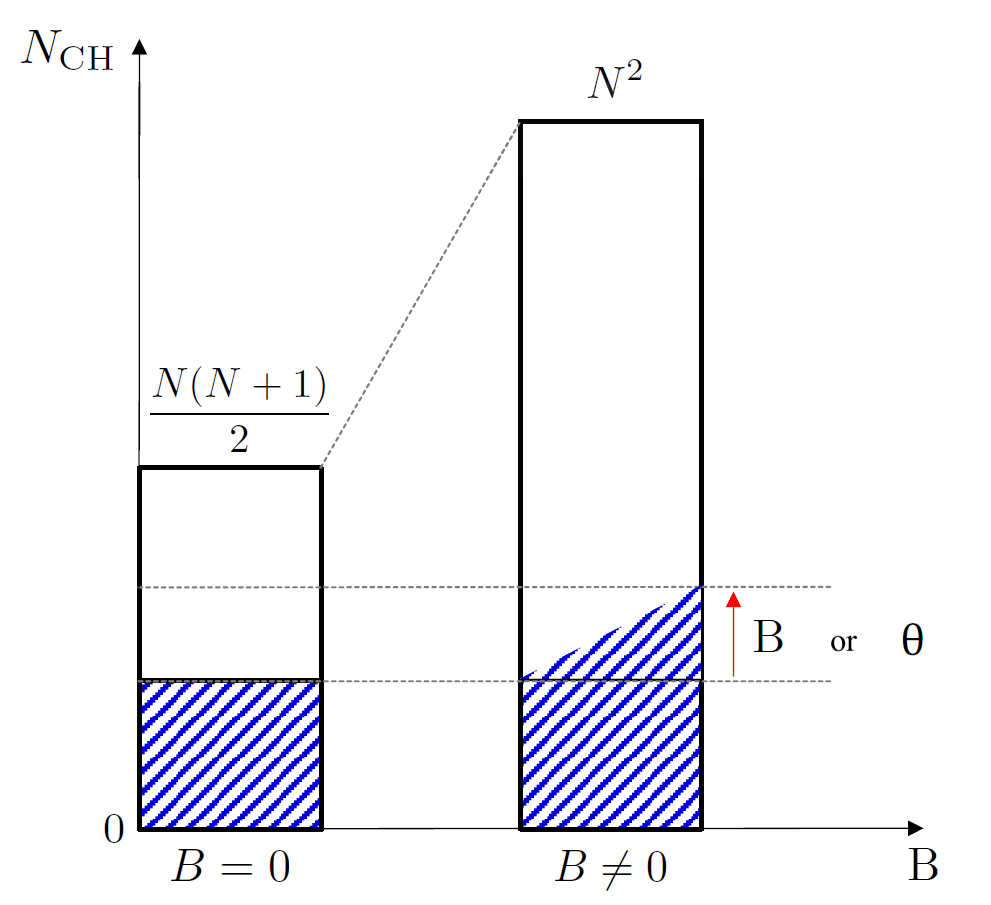}
    \caption{}
\end{subfigure}
\begin{subfigure}{0.45\textwidth}
    \includegraphics[width=1\textwidth]{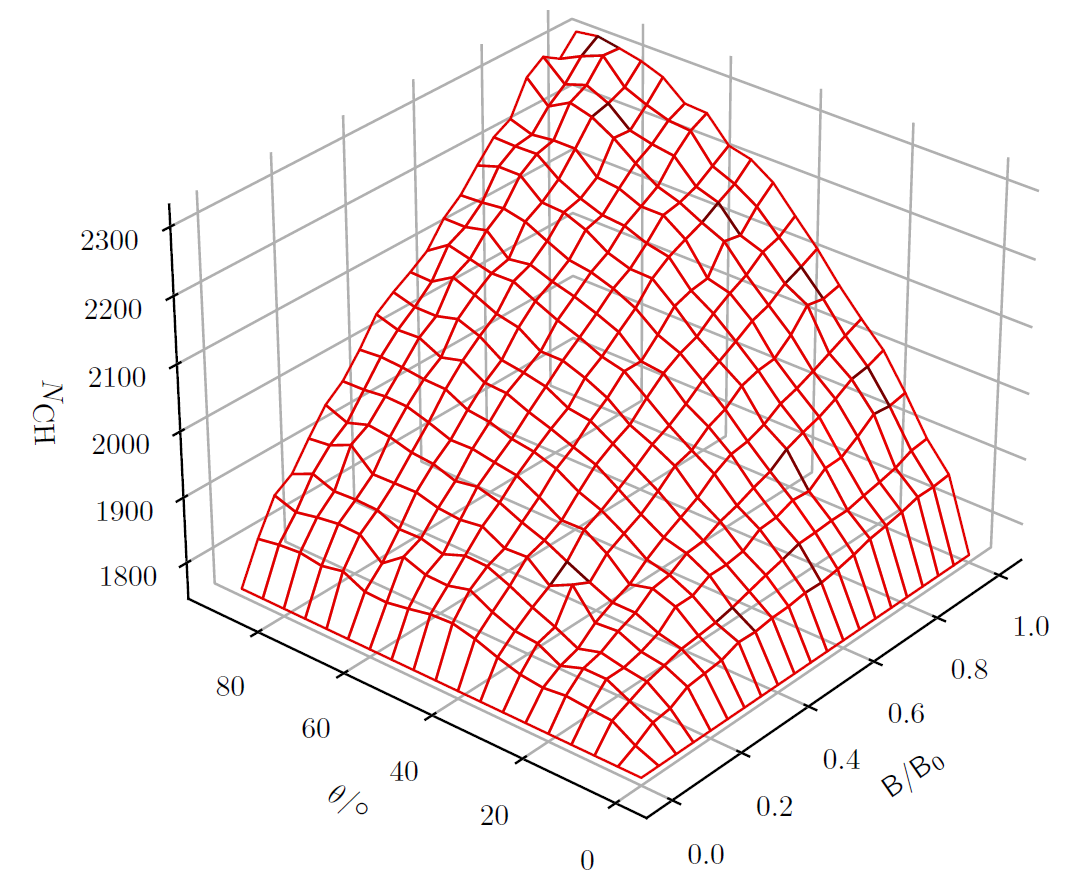}
    \caption{}
\end{subfigure}
    \caption{Number of Cholesky vectors as a function of the magnetic-field strength $B$ and orientation $\theta$ for an energy calculation of the CH molecule at the ff-MP2 level of theory with the unc-aug-cc-pVQZ basis set using a CD of the ERIs.\\ (a) Qualitative depiction. Reprinted from ref.~\citenum{Blaschke24c}. \\ 
    (b) Calculated number of Cholesky vectors for $\delta=9$. Reprinted from Blaschke, S.; Stopkowicz, S. CD of complex valued ERIs over GIAOs: Efficient MP2 computations for large molecules in strong magnetic fields. \textit{J. Chem. Phys.} \textbf{2021}, 156, 044115, 
    with the permission of AIP Publishing.\cite{Blaschke22} }
    \label{fig:CH_ffMP2}
\end{figure}

\textbf{The methylidyne radical in a magnetic field}:   
In order to study the dependence of the compression rate of the CD to the magnetic-field strength and orientation, the methylidyne radical (CH) was examined in Ref.~\onlinecite{Blaschke22}. The number of Cholesky vectors as a function of the magnetic field for an ff-CD-MP2 calculation using an unc-aug-cc-pVQZ basis\cite{Dunning89,Kendall92} computed within \cfour\ can be found in Fig.~\ref{fig:CH_ffMP2}. The magnetic-field strength is $B=0\textrm{-}1.0\ B_0$ for angles between $\theta=0^\circ$ (parallel orientation) and $\theta = 90^\circ$ (perpendicular orientation). Results are shown for a Cholesky threshold $\tau$ of $10^{-9}$. It has to be noted that for linear molecules in a parallel orientation, the phase factor of the GIAOs is one and thus the GIAOs, unlike for the general case, do not depend on the magnetic field. This is due the fact that the cross product of parallel vectors is zero and, as such, the exponent of the phase of the London orbital vanishes  
\begin{equation}
	{\bf{B}}\cross({\bf{R}}_O-{\bf{R}}_\mu)=0 \Leftrightarrow  {\bf{B}} \parallel  ({\bf{R}}_O-\bf{R}_\mu). \label{eq:ERIspar}
\end{equation}
In this case, the eightfold symmetry is restored and the ERIs are real. For all other orientations, the permutational symmetry is reduced and the ERIs are complex valued.
The minimum number of Cholesky vectors for a given threshold is generated for $B=0 \ B_0$ and for $\theta = 0^\circ$.
The Cholesky rank increases almost linearly with both increasing $B$ and $\theta$ reaching a maximum for $B=1\ B_0$ in the perpendicular orientation $\theta = 90^\circ$. 
Given the fact that the CD can be seen as a modified Gram-Schmidt orthogonalization of the product densities $| \sigma \rho)$ that eliminates linear dependencies of the basis set up the to Cholesky threshold, the following is inferred. 
The linear dependencies decrease for an increasing magnetic-field strength.
Moreover, starting from a fully restored eightfold permutational symmetry in the parallel case, the linear dependencies decrease when the molecule is rotated towards a perpendicular magnetic-field orientation. 
As such, the maximum compression rate is observed for the parallel case. Since the ERIs in the parallel case are constant for different magnetic-field strengths (see Eq.~\eqref{eq:ERIspar}), the compression rate is field-strength independent, too.\cite{Blaschke22} 
This observed trend demonstrates how the CD successfully tempers the increased computational cost that arises from the reduction of the eightfold permutational symmetry of the ERIs to fourfold in the ff case.

\textbf{Mg transition pressure broadening in a He atmosphere}:
Recently, absorption lines in the spectrum of a strongly magnetic white dwarf star (SDSS J1143+6615) have been assigned to metal atoms including Na and Mg for the first time.\cite{Hollands23} 
Specifically, the Mg absorption line which arises from the electronic excitation $3p \rightarrow 4s$ $({^3P}_u \rightarrow {^3S}_g$) proved very important. In the presence of a magnetic field, the $p$ orbitals split to their corresponding $m_l=-1,0,+1$ components. While a perturbative consideration that accounts only for the orbital-Zeeman interaction predicts the transition from the $3p_0$ orbital to be constant in different magnetic field strengths, full ff calculations predict a blue shift for this component.\cite{Hollands23,Kitsaras24} This deviation was vital for the characterization of the spectrum in question.\cite{Hollands23}
Furthermore, in non-magnetic white dwarfs this absorption shows an asymmetric feature which is a pressure broadening effect originating from the interaction of the Mg atom with the He dominated atmosphere.\cite{Allard16} This pressure broadening was studied at the ff CC level of theory using Cholesky decomposition in Ref.~\onlinecite{Blaschke24b}.

 \begin{figure}[H]
    \centering
        \includegraphics[width=0.4\textwidth]{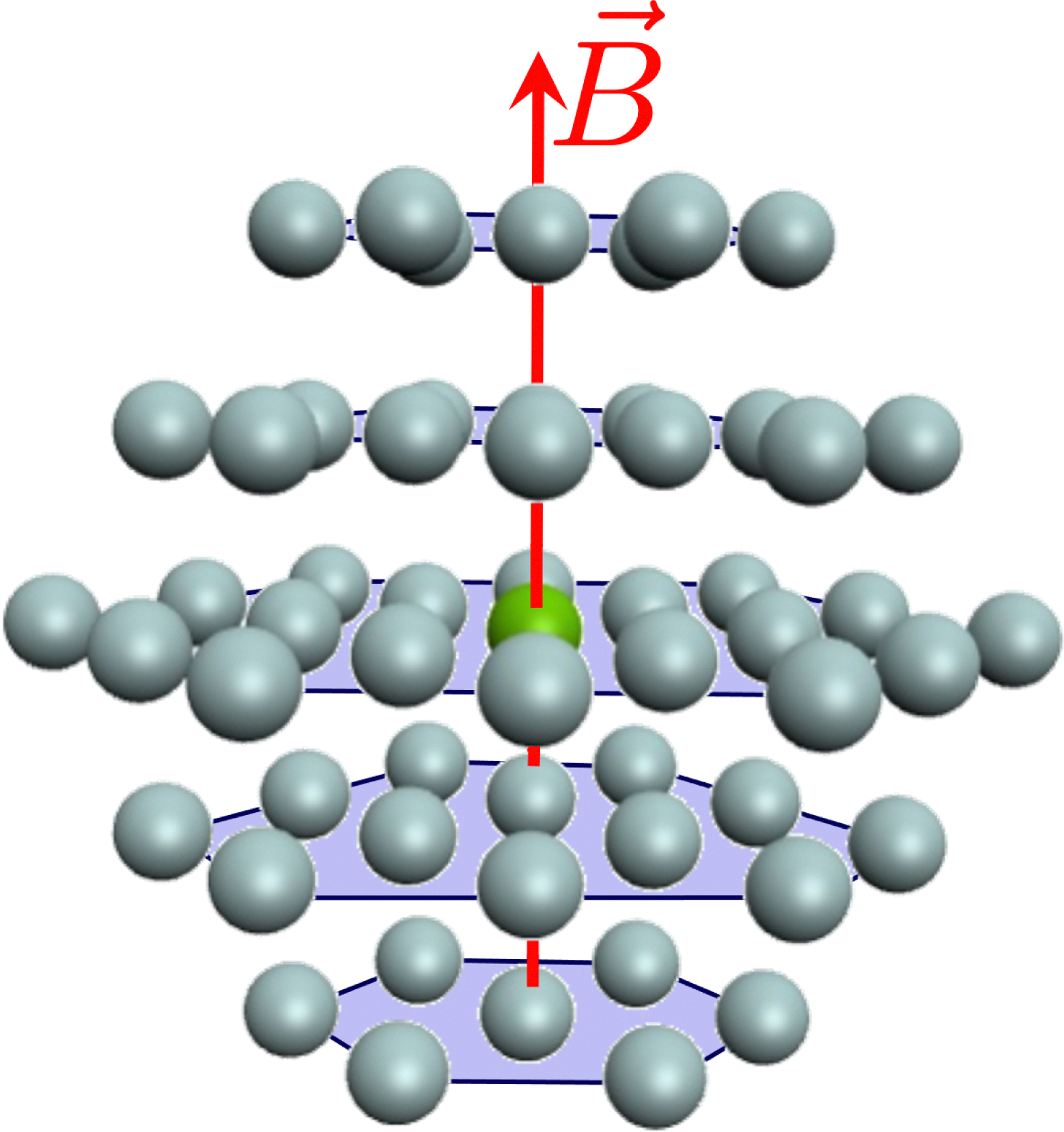}
    \caption{Reproduced from Blaschke, S.; Kitsaras, M.-P.; Stopkowicz, S. Finite-field Cholesky decomposed coupled-cluster techniques (ff-CD-CC): theory and application to pressure broadening of Mg by a He atmosphere and a strong magnetic field. \textit{Phys. Chem. Chem. Phys.} \textbf{2024}, 26,28828–28848, licensed under CC BY 3.0.\cite{Blaschke24b}\\
    Representation of a Mg atom (green) in an hcp lattice of He atoms (gray). Depicted are the first two shells forming a cluster of \ce{MgHe56}, where 12 He atoms are in the first and 44 He atoms in the second shell.}
    \label{fig:MgHen}
\end{figure}
An explicit solvation model was used to simulate the effects of the He atmosphere on the Mg transition in question. 
Beyond studying the Mg-He dimer in a magnetic field, MgHe$_n$ clusters as depicted in fig.~\ref{fig:MgHen} were examined at the ff-EOM-CC2 and ff-EOM-CCSD levels of theory, with $n=12$ for the first coordination sphere in a hexagonal closed-packed structure and $n=56$ when considering the second shell as well.\cite{Blaschke24b,Blaschke24c} The unc-aug-cc-pCVQZ-basis set was used for Mg and the unc-aug-cc-pVDZ basis set for the He atoms. The He density $\rho$ is probed by varying the distance of the He atoms around the Mg center.
\cfour\ was used for the generation of Cholesky vectors and a ff-SCF solution while \qcumbre\ was employed for the post-HF calculations. 
The use of CD in the case of the two-shell cluster reduced the memory requirements for storing the ERIs from $6 \ \textrm{TB}$ to $34 \ \textrm{GB}$ with a Cholesky threshold of $10^{-5}\ E_\mathrm{h}$.
Hence, these calculations were only feasible by employing CD techniques.\cite{Blaschke24b}

 \begin{figure}[H]
    \centering
        \includegraphics[width=0.6\textwidth]{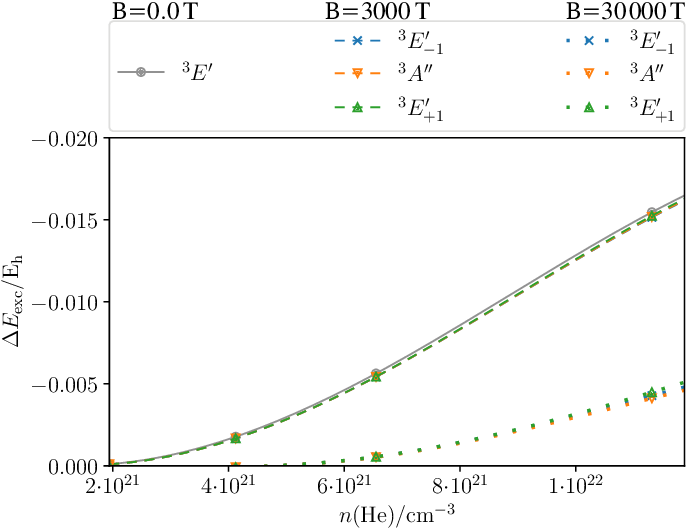}
    \caption{Reproduced from Blaschke, S.; Kitsaras, M.-P.; Stopkowicz, S. Finite-field Cholesky decomposed coupled-cluster techniques (ff-CD-CC): theory and application to pressure broadening of Mg by a He atmosphere and a strong magnetic field. \textit{Phys. Chem. Chem. Phys.} \textbf{2024}, 26,28828–28848, licensed under CC BY 3.0.\cite{Blaschke24b}\\
    Shift in the excitation energy $\Delta E_\mathrm{exc} = E_\mathrm{exc}(\rho \rightarrow 0) - E_\mathrm{exc} (\rho)$ (in Hartree) at the CD-CCSD ($\tau = 5$) level of the Mg triplet transition as a function of the helium density $\rho$ of the \ce{MgHe12} ($C_{3h}$) atmospheric model system in the field-free case as well as in a magnetic field of $3\ 000 \ \textrm{T}$ and $30\ 000 \ \textrm{T}$ oriented along the $C_3$ axis.
    Employing the unc-aug-cc-pCVQZ on Mg and the unc-aug-cc-pVDZ basis on He.}

    \label{fig:Mgshift}
\end{figure}
In Fig.~\ref{fig:Mgshift}, the shift of the transition energy is plotted as a function of the He density $\rho$ for the field-free case, for $B=3\ 000 \ \textrm{T}$ and for $B=30\ 000 \ \textrm{T}$.
The study has successfully replicated the prediction for the asymmetric feature of the Mg absorption in the absence of a magnetic field due to the effects of the He atmosphere. 
This follows from predicting a blue shift for the transition in question in increasing He density. For a magnetic field strength of around $3\ 000 \ \textrm{T}$, which is the field strength assigned to SDSS J1143+6615 discussed above, the effects of the He atmosphere are very similar to the field-free case.
For this reason, it is deemed an appropriate approximation to model the line in the same manner as for white dwarfs that exhibit no or very low magnetic fields. 
However, in stronger magnetic fields around  $30\ 000 \ \textrm{T}$, the overall pressure effects are significant only for very high densities. 
It can thus be concluded that the pressure-broadening effects are not expected to be prominent on the shape of the absorption line for the studied transition in very strong magnetic fields. For further details on this study, the reader is referred to the original publication.\cite{Blaschke24b}\\
Current developments considering CD together with finite-field calculations focus on the treatment of relativistic effects as well as dynamics.


\section{Conclusions and outlook}

We provided in this paper a 
status report on the use of the Cholesky decomposition of the two-electron integrals within the \cfour\ quantum-chemical program package. The available features are summarized in Table~\ref{tab:summary}
and will be included in the next release of \cfour\ planned for 2027.
\begin{table}
    \centering
    \begin{tabular}{lccccc}
        \toprule
        Method & Energy & Forces & NMR shieldings & Magnetizability & FD properties  \\
        \midrule
        HF & \checkmark & \checkmark & \checkmark & \checkmark & \checkmark$^*$ \\
        MP2 & \checkmark & \checkmark & X & X & X \\
        CASSCF & \checkmark & \checkmark & \checkmark & X & \checkmark \\
        CCSD & \checkmark & \checkmark & X & X & X\\
        \bottomrule
    \end{tabular}
    \caption{Summary of the present capabilities of \cfour\ that can be accelerated using Cholesky Decomposition. $^*$HF linear response and frequency dependent properties are available as a special case of CASSCF}
    \label{tab:summary}
\end{table}
Among the available features, we highlight in particular
(a) the use of symmetry in the CD of ERIs and their derivatives, (b) the use of CD in the treatment of magnetic properties (in particular NMR shieldings) 
when using GIAOs to ensure gauge-origin independence, and (c) the use of CD within ff methods to treat atoms and molecules in the presence of strong magnetic fields. Other developments, including the use of CD in CASSCF and CC computations, 
analytical evaluation of geometrical derivatives, and relativistic treatments further extend the applicability of the corresponding methods, although related  developments have also been reported by others. The examples given in this review demonstrate how CD extends the applicability of electron-correlated methods to larger systems. 

Future work in \cfour\ concerning CD will address the following issues. First, further improvements in the efficiency of the generation of the Cholesky vectors are desirable, in particularly for what concerns the evaluation of the underlying two-electron integrals. Recent developments for accelerating integral evaluation within the McMurchie-Davidson scheme, such as the ones reported in Ref.~\onlinecite{Neese23}, may provide a promising route in this direction.

A natural next step for the CD-based CC developments is the inclusion of higher excitations. In particular, a CD based CCSD(T)\cite{Raghavachari89} scheme is 
essential to extend the present framework to chemical accuracy, which is highly desirable in a large manifold of applications. Likewise, our CD based CASSCF scheme needs to be extended 
to account for dynamical correlation on top of the CASSCF treatment. Among the various options, the most popular ones are the ones based on a second-order perturbation approach such as CASPT2\cite{Andersson1990,Andersson1992} or NEVPT2.\cite{Angeli2001,Angeli2001a,Angeli2002} A particularly attractive alternative is represented by the Adiabatic Connection formalism, recently developed and introduced by Pernal and coworkers,\cite{Pernal18,Pastorczak18,Drwal22} as it represents a cheaper, yet accurate alternative to the perturbative methods. 
Concerning the evaluation of properties, on-going work focuses on the extension of the corresponding schemes to the CC level, to make magnetizability computations available at MP2 and CASSCF, investigate the possibility to compute vibrational frequencies (numerically or analytically) using CD, 
provide electronic excitation energies (and also ionization potentials and electron affinities) within
EOM-CC/LR-CC treatments and to tackle the issue of property computation within relativistic methods, where current work (except for first-order properties at the X2C-CCSD\cite{Zhang25}) so far was only concerned with energy evaluations.

A further important issue is that CD does not 
reduce the scaling of CC calculations and instead only decreases prefactors and reduces both memory and I/O requirements. Thus, to extend the applicability of CC methods further, it is essential 
to combine the CD based CC schemes with local correlation methods.\cite{Saebo89,Hampel96,Schuetz99b,Hetzer00,Schuetz00b,Schuetz01,Li01,Li06,Rolik11,Rolik13,Riplinger13,Riplinger13b,Ziolkowski10,Eriksen15b,Fishman26}
A challenge is here to identify an ansatz that remains computationally so simple that the evaluation of properties, one of the strengths of \cfour, remains possible. A recent implementation of DLPNO-MP2 second derivatives\cite{Stoychev21} shows how complicated things can become otherwise.

In conclusion, the developments summarized in this contribution provide a flexible and efficient foundation for the future extensions mentioned here, and demonstrate that Cholesky decomposition can serve as a common computational framework for extending \cfour\ capabilities as a high-accuracy code for molecular energies, structure, and properties to larger molecular systems. 

\section*{Acknowledgement}

In Mainz (J.G.) and Saarbr{\"u}cken (S.S.), the work on the use of Cholesky decomposition in \cfour\ was supported by the Deutsche Forschungsgemeinschaft (DFG) in the framework of the collaborative research center "Multiscale Simulation Methods for Soft-Matter Systems" (TRR 146) under Project No. 233630050. Funding for the work in Saarbr{\"u}cken was also provided by the Deutsche Forschungsgemeinschaft via grant STO-1239/1-1 and  grant 510228793/SFB1633 (TP A01). The work at Johns Hopkins University (L. C.) has been supported by the National Science Foundation, under Grant No. PHY2309253. 
The work in Pisa (F. L.) has been supported by the Italian Ministry of University and Research (PRIN 2022) under grant 2022WZ8LME$\_$002 and from ICSC -- Centro
Nazionale di Ricerca in High Performance Computing,
Big Data, and Quantum Computing, funded by the European
Union-NextGenerationEU-PNRR, Missione 4 Componente 2
Investimento 1.4.
\section*{Data availability}
The data that support the findings of this article are available within the article

\bibliography{bibliography}

\end{document}